%% file: DiffuseGalacticEmission.tex
\documentclass[useAMS,usenatbib,a4paper]{mn2e}
\usepackage{color}
\usepackage{graphicx}
\usepackage{amssymb}
\usepackage{amsmath}
\usepackage{fixltx2e}
\usepackage[breaklinks, colorlinks, citecolor=blue]{hyperref}
\usepackage{epstopdf}

\providecommand{\sorthelp}[1]{}
\definecolor{DarkGreen}{rgb}{0.0, 0.5, 0.0}
\definecolor{purple}{rgb}{0.6, 0.0, 0.4}


\title[C-Band All-Sky Survey: A First Look at the Galaxy]{C-Band All-Sky Survey: A First Look at the Galaxy}
\author[M.\,O.\,Irfan et al.]{M.\,O.\,Irfan,$\!^{1}$\thanks{\url{E-mail: mirfan@jb.man.ac.uk}} C.\,Dickinson,$\!^{1}$\thanks{\url{E-mail: Clive.Dickinson@manchester.ac.uk}} R.\,D.\,Davies,$\!^{1}$ C.\,Copley,$\!^{2,3,4}$ R.\,J.\,Davis,$\!^{1}$ P.\,G.\,Ferreira,$\!^{4}$
\newauthor C.\,M.\,Holler,$\!^{4,5}$ J.\,L.\,Jonas,$\!^{2,3}$ Michael\,E.\,Jones,$\!^{4}$ O.\,G.\,King,$\!^{4,6}$ J.\,P.\,Leahy,$\!^{1}$ J.\,Leech,$\!^{4}$ 
\newauthor  E.\,M.\,Leitch,$\!^{7}$ S.\,J.\,C.\,Muchovej,$\!^{6}$ T.\,J.\,Pearson,$\!^{6}$ M.\,W.\,Peel,$\!^{1}$ A.\,C.\,S.\,Readhead,$\!^{6}$
\newauthor M.\,A.\,Stevenson,$\!^{6}$ D.\,Sutton,$\!^{4,8,9}$  Angela\,C.\,Taylor,$\!^{4}$ J.\,Zuntz $\!^{1,4}$  \\
$^{1}$Jodrell Bank Centre for Astrophysics, School of Physics and Astronomy, The University of Manchester, Oxford Road \\
Manchester, M13 9PL, Manchester, U.K. \\
$^{2}$Department of Physics and Electronics, Rhodes University, Drostdy Road, Grahamstown, 6139, South Africa \\
$^{3}$SKA SA, 3rd Floor, The Park, Park Road, Pinelands, 7405, South Africa \\
$^{4}$Sub-department of Astrophysics, University of Oxford, Denys Wilkinson Building, Keble Road, Oxford OX1 3RH, U.K. \\
$^{5}$Munich University of Applied Sciences, Lothstr. 34, 80335 Munich, Germany \\
$^{6}$California Institute of Technology, Pasadena, CA 91125, USA \\
$^{7}$University of Chicago, Chicago, Illinois 60637, USA \\
$^{8}$Institute of Astronomy, University of Cambridge, Madingley Road, Cambridge CB3 0HA, U.K. \\
$^{9}$Kavli Institute for Cosmology, Cambridge, Madingley Road, Cambridge, CB3 0HA, U.K.
}

\begin{document}


\pagerange{\pageref{firstpage}--\pageref{lastpage}} \pubyear{2014}

\maketitle

\label{firstpage}

\begin{abstract}
We present an analysis of the diffuse emission at 5\;GHz in the first
quadrant of the Galactic plane  using two months of preliminary intensity data
taken with the C-Band All Sky Survey (C-BASS) northern instrument at the Owens Valley Radio Observatory,
California.

Combining C-BASS maps with ancillary data to make temperature-temperature plots we find
synchrotron spectral indices of $\beta = -2.65 \pm 0.05$ between 0.408\;GHz 
and 5\;GHz and $ \beta = -2.72 \pm 0.09$ between 1.420\;GHz and
5\;GHz for $-10\degr < |b| < -4\degr$, $20\degr < l <
40\degr$. Through the subtraction of a radio recombination line
(RRL) free-free template we determine the synchrotron spectral index
in the Galactic plane ($ |b| < 4\degr$) to be $\beta = -2.56
\pm 0.07$ between 0.408\;GHz and 5\;GHz, with a contribution of $53 \pm 8$ per cent from free-free emission at 5\,GHz. These results
are consistent with previous low frequency measurements in the
Galactic plane.

By including C-BASS data in spectral fits we demonstrate the presence
of anomalous microwave emission (AME) associated with the {H\sc{ii}}
complexes W43, W44 and W47 near 30\;GHz, at 4.4$\,\sigma$,
3.1$\,\sigma$ and 2.5$\,\sigma$ respectively. The CORNISH VLA 5\,GHz
source catalogue rules out the possibility that the excess emission
detected around 30\;GHz may be due to ultra-compact H{\sc ii}
regions. Diffuse AME was also identified at a 4$\sigma$ level within
$30\degr < l < 40\degr$, $-2\degr < b < 2\degr$ between 5\;GHz and
22.8\;GHz.

\end{abstract}


\section{Introduction}

Diffuse Galactic radio emission is a combination of free-free,
synchrotron, anomalous microwave and thermal dust emission. Below 10\;GHz,
free-free and synchrotron contributions account for the majority
of the emission, while anomalous microwave emission (AME) reaches
its peak between 10 and 30\;GHz \protect\citep{Gold2009,PlanckAME13}. At frequencies $
>$ 100\;GHz, the total Galactic emission is dominated by thermal
emission from dust {\protect\citep{FinkDust}}. Measurements of the diffuse
Galactic emission are used for both the interpretation of cosmological
data and the understanding of our Galaxy. Synchrotron emission can
reveal information on the Galactic magnetic field and cosmic ray
physics {\protect\citep{Jaffe}} while free-free emission measurements can be 
used to explore
{H\sc{ii}} regions and the warm ionised medium {\protect\citep{D96}}. 
The cosmological
demand for increasingly sensitive, large-scale maps of Galactic
emission is driven by the need to remove these
`foreground' emissions to obtain an accurate measurement of the
cosmic microwave background (CMB) signal. This is particularly the case for imaging the CMB polarization signal and identifying $B$-modes, where the intrinsic cosmological signal is a small fraction of the foregrounds \protect\citep{Dunkley2009,Betoule2009,Wmap09,Errard2012}. This issue has been exemplified by the recent claim of a detection of intrinsic CMB $B$-modes \protect\citep{Bicep2014}, which now appears to be contaminated by foregrounds \protect\citep{Planck2014_Int_XXX}. 

As different foregrounds dominate over different spectral and spatial
ranges, it is important to have all-sky data
available covering a range of frequencies and with $\sim 1\degr$
resolution if CMB polarization experiments are to
live up to their full potential in determining cosmological parameters
{\protect\citep{CompSep13}}. The low frequency all-sky intensity surveys 
currently most used for this purpose are at
0.408\,GHz {\protect\citep{Haslam}}, 1.420\,GHz {\protect\citep{Reich, Testori,
    Reich01}}, \mbox{22.8\,GHz} {\protect\citep{Wmap09}} and
 28.4\,GHz {\protect\citep{Planck13}}. {\protect\citet{Jonas}} and {\protect\citet{spass}}
provide southern Hemisphere maps at 2.3\,GHz in intensity and
polarization, respectively. {\protect\citet{sino}} present a polarization
survey of the Galactic plane at 5\,GHz with a FWHM 
of \mbox{9.5 arcmin} so as to capture the small-scale structures. 
However the lack of all-sky, degree scale surveys between 1.420\,GHz and 22.8\,GHz introduces great
uncertainties in the spectral behaviour of free-free,
synchrotron and anomalous microwave emission across the full sky \protect\citep{PeelSync}.

C-BASS, the C-Band All Sky Survey, is a project currently mapping
Galactic intensity and linear polarization at a frequency of 5\;GHz
(Jones et al. in prep, {\protect\citealt{King10}}). At this frequency
the polarized emission is negligibly affected by Faraday rotation at
high latitudes ($\Delta\psi\lesssim 1^{\circ}$ across most of the $ |b|> 10\deg$ latitude sky). C-BASS consists of two independent
  telescopes: \mbox{C-BASS} North observes from the Owens Valley
Radio Observatory (OVRO) in California, USA (latitude $37\degr\!\!.2$
N) whilst C-BASS South is located at the Klerefontein support base for
the Karoo Radio Astronomy Observatory (latitude $30\degr\!\!.7$
S). C-BASS North achieved first light in 2010 and entered its final
survey mode in late 2012 after a period of commissioning and upgrades
(Muchovej et al. in prep). C-BASS South is currently being
commissioned. The final all-sky maps will have a FWHM resolution of
\mbox{43\farcm 8} and a nominal sensitivity in polarization of {$\approx$ 0.1\;mK per beam}.
On completion the C-BASS data
will be used to probe the Galactic magnetic field using synchrotron
radiation, constrain the spectral behaviour of AME, and provide a
polarized foreground template for CMB polarization experiments. The
large angular scale and high sensitivity of the final survey will be
of particular use for confirming $B$-mode detections at low
multipoles.

In this paper we present some of the Galactic science that can be
achieved using preliminary C-BASS intensity results. We focus on the
composition of the Galactic plane at 5\;GHz, determining the spectral
index of synchrotron emission between 0.408 and 5\;GHz and the
fractional composition of free-free and synchrotron emission
present. We extend this analysis to include higher frequency data to
constrain the AME amplitude and spectrum towards a few compact
regions. For this analysis we use only two months of preliminary
C-BASS North intensity data (January and February 2012), which has a sufficient signal-to-noise ratio to constrain the properties of the bright emission near the Galactic plane. The main observational parameters for 
C-BASS North are shown in Table~\ref{tab:obs}.

\begin{table}
\caption{C-BASS North observational parameters. 
The full-beam area is defined as being within 3\fdg 5 of the main 
beam peak and the first main-beam null is at 
1\degr\ from the main-beam peak. \label{tab:obs}}
 \begin{tabular}{ l  c }
\hline
Parameter & C-BASS North \\
\hline \hline
Latitude & 37\fdg 2 N  \\
Antenna optics & symmetric Gregorian \\
Antenna geometry & Az/El \\
Diameter & 6.1\,m  \\
Primary beamwidth (FWHM) & $44$\,arcmin \\
First Null & $1\fdg5$ \\
Main-beam efficiency ($<1\fdg 0$) & 72.8\,$\%$ \\
Full-beam efficiency ($<3\fdg 5$) & 89.0\,$\%$ \\
Intensity centre frequency & 4.76\,GHz\\
Intensity noise-equivalent bandwidth  & 0.489\,GHz \\
Noise equivalent temperature & 2\,mK$\sqrt{\rm{s}}$ \\
\hline
\end{tabular}
\end{table}

This paper is organized as follows: Section 2 describes the C-BASS and
ancillary data, and presents a derivation of the free-free template
used in subsequent analysis.  The quality of the C-BASS data
calibration is verified in Section 3.  In Section 4 we investigate the
synchrotron spectral index of the Galactic plane ($|b| < 4\deg$), 
determining the
intermediate latitude ($4\degr < |b| < 10\degr$) synchrotron-dominated
spectral index between 0.408/1.420\,GHz and 5\;GHz, as well as the
in-plane ($0\degr < |b| < 4\degr$) synchrotron spectral index once we
account for free-free emission in this region. In Section 5 we use the C-BASS data alongside higher-frequency data from
the \emph{WMAP}  and \emph{Planck} satellites
 to constrain the spectral behaviour of anomalous
microwave emission in $-2\degr\!\! < b < 2\degr\!\!$,
$30\degr < l < 40\degr$. Conclusions are presented in Section 6.

\section{Data}

\subsection{C-BASS}

C-BASS North observes by repeatedly scanning over
  360\degr\ in azimuth at a constant elevation. Several different scan
  speeds close to 4\degr\ per second are used, to ensure that a given
  feature in the time-ordered data does not always
  correspond to the same angular frequency on the sky, in case of data corruption by effects such as microphonics. The C-BASS North data are subject to 1.2\;Hz (and harmonics) microphonic oscillations but these effects are reduced to a negligible level within the intensity data, especially within the high signal-to-noise regions investigated here, by our data reduction pipeline (Muchovej et al., in prep.). We scan at an
  elevation of 37\fdg 2 (through the North Celestial Pole) for about
  two-thirds of the time, and at higher elevations (mostly 47\fdg 2)
  for the remainder of the time, in order to even out the sky coverage. The
northern receiver is a continuous-comparison receiver
{\protect\citep{King}}, which measures the difference between sky
brightness temperature and a temperature-stabilized resistive
load. This architecture reduces receiver 1/$f$ noise, which would
otherwise contaminate the sky signal.

The northern receiver has a nominal bandpass of \mbox{4.5 --
  5.5\;GHz}, but in-band filters to remove terrestrial radio frequency
interference (RFI) reduce the effective bandwidth to 0.489\;GHz, with
a central frequency of 4.76\;GHz {\protect \citep{King}}. The finite and relatively large (20 per cent) bandwidth, means that adopting a single central frequency is a simplification. This is because the effective frequency is a convolution of the instrument bandpass with the sky signal across this bandpass:
\begin{equation}
\nu_{\rm{eff}}^{\beta} = \frac{\int f(\nu) G_{m} (\nu) \nu^{\beta} d\nu}{\int f (\nu) G_{m} (\nu) d\nu},
\end{equation}
where $f(\nu)$ is the bandpass response and $G_{m}(\nu)$ is the forward gain {\protect \citep[e.g.,][]{Jar11}}. Any source that has a spectral shape different to that of the calibrator source, will result in a slightly different effective frequency. One can correct for this by making `colour corrections' {\protect \citep[e.g.,][]{CC}} assuming the bandpass shape and the spectral shape of the source being observed. For C-BASS, we estimate that these colour corrections are $\approx$ 1 per cent for typical spectra \citep{IrfanThesis}; only spectral shapes significantly different to that of our calibrators (primarily Tau A and Cas A) require corrections. For this analysis, we do not make any colour corrections and assume an effective frequency of 4.76\;GHz. We include an additional 1 per cent uncertainty, which is added in quadrature with the other uncertainties. 

The raw telescope time-ordered data are processed by a data
reduction pipeline that identifies RFI events, performs amplitude and
polarization calibration and applies atmospheric opacity corrections
(Muchovej et al. in prep). The northern receiver includes a noise
diode which, when activated, injects a signal of constant noise
temperature. The noise diode excess noise ratio (ENR) is stable to within \mbox{1 per cent} over time periods of several months. The data reduction pipeline calibrates the intensity
signal onto the noise diode scale, and the noise diode is subsequently
calibrated against the astronomical sources Cas A, Tau A and Cyg
A. The calibrator point-source flux densities are calculated from the
spectral forms given in {\protect\citet{Weiland}},
{\protect\cite{Baars}} and {\protect\citet{Vinyaikin}} and converted
to an equivalent antenna temperature. Next, the models of the beam are used to convert the antenna temperature scale ($T_{A}$) to a `full beam' temperature scale ($T_{FB}$):
\begin{equation}
T_{FB} = T_{OFF} + \frac{\Omega_{A}}{\Omega_{FB}} T_{A},
\end{equation}
where $\Omega_{FB}$ is the full beam solid angle, $\Omega_{A}$ is the total beam solid angle and $T_{OFF}$ is the survey zero level {\protect \citep[e.g.,][]{ Jonas}. Following \protect\cite{ReichI} and \protect\cite{Jonas}, we
define the extent of the full-beam to be 3\fdg 5 from boresight. Using
GRASP physical optics simulations {\protect\citep{Holler}} we
calculate the C-BASS North antenna temperature to full-beam brightness temperature
conversion factor to be 1.124. The initial temperature scale of the C-BASS maps is referenced to the antenna
temperature of the noise diode, and this is subsequently converted antenna temperature and then
the more useful full-beam brightness temperature to account for power
lost in the far sidelobes. The level of power lost is dependent on the
observed source morphology and so this factor varies depending on what
is being observed. It is the generally accepted practice to convert
from antenna temperature to the full-beam scale, as described above,
in order to fully represent the large-scale structures of the
map. While using $T_{FB}$ is correct for diffuse emission, as demonstrated in Section ~\ref{sec:dataVal}, it underestimates point source emission by $\approx 5$ per cent and so $T_{A}$ is used for the analysis conducted in Section ~\ref{sec:sed}.   

The conversion between flux density and full-beam brightness temperature
required for this calibration scheme, alongside the opacity (calculated
from skydips) and colour
corrections, carry with them a few per cent uncertainty. The main
source of uncertainty in the calibration comes from the telescope 
aperture efficiency, which is required to convert between flux density
and antenna temperature. The aperture efficiency of the northern telescope
has been determined by simulations, to an accuracy that we estimate to
be 5 per cent, dominated by uncertainties in the modeling of standing 
waves between the feed and the subreflector. Further work is being done
on both simulation and measurement of the aperture efficiency for the 
calibration of the full survey. In the meantime a conservative 5
per cent calibration error has been assigned to these preliminary data.

 The C-BASS maps are constructed from the
  pipeline-processed time-ordered data using the DEStriping
  CARTographer, \textsc{Descart}
  {\protect\citep{Descart}}. \textsc{Descart} performs a maximum
  likelihood fit to the data, by modelling the contribution of
  1/\emph{f} noise in the timestreams as a series of offsets of a
  given length. The multiple crossings of each pixel over a range of
  parallactic angles in long sets of observations allow these offsets
  to be well determined. The characteristic timescale (`knee
  frequency') of typical $1/f$ fluctuations in the C-BASS intensity
  data is tens of mHz (corresponding to tens of seconds), depending on
  the atmospheric conditions. We adopt an offset length of 5\,seconds,
  although we achieve similar results with 10\,seconds. We use power
  spectra of the time-ordered data to estimate the white-noise
  variance and knee frequency, providing a model of instrumental noise
  which is used in the mapping process. The data are inverse-variance
  weighted in each pixel to achieve the optimal weighting for
  signal-to-noise ratio. No striations are visible in our
  final map. The data are gridded into {\tt{HEALPix}}
  {\protect\citep{Gorski}} maps with $N_{\rm side} = 256$,
  corresponding to 13.7 arcmin pixels.

\begin{figure}
\centering
\includegraphics[width=0.5\textwidth]{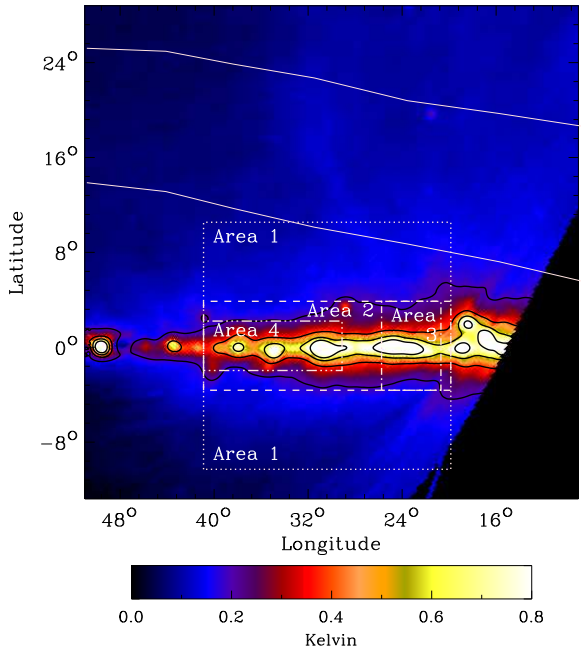} 
\caption{C-BASS 4.76\;GHz intensity map using data taken in January
  and February 2012 showing a region in the first Galactic
  quadrant. The data are at $N_{\rm side} = 256$ and FWHM of
  44\,arcmin. The white dashed rectangles highlight the four regions, 
  used in this analysis 
  and the white solid curved lines delineate the northern part of the
  Gould Belt. The empty map region, bottom right, is due to an absence
  of C-BASS North data below declination = $-30\degr$; this area will
  be mapped by C-BASS South observations in the future. The colour bar
  scale is in kelvin and the black contours are spaced at 0.2\;K
  intervals.}
\label{fig:Area}
\end{figure}

The completed intensity survey made using all available data will be
confusion limited, but the maps used in this analysis were made with
just two months data and are noise limited. The C-BASS
  confusion noise limit in intensity is estimated to be $\approx
  130$\,mJy\,beam$^{-1}$ \citep{conNoise} or 0.8\,mK, while the
  thermal noise limit for the two month preliminary data used in this
  analysis is $\approx 6$\,mK. The Galactic signals analysed in this
  paper are $>$ 500\,mK.


The four regions listed in Table~\ref{tab:areas} and outlined
in Fig.~\ref{fig:Area}. The four regions
can be considered as off-plane (Area 1, which is split into two) and in-plane (Areas 2, 3 and
3). Fig.~\ref{fig:Area} shows the Galactic plane and higher
latitudes as seen at 4.76\;GHz by \mbox{C-BASS} North as well as the northern part of the Gould Belt, a star forming disc tilted at 18\degr\ to the Galactic plane. 
The Gould Belt is known for its increased supernova activity {\protect\citep{Grenier}}
and includes the Ophiuchus cloud complex.

\begin{table}
\setlength{\tabcolsep}{1mm}
\caption{The four Galactic plane regions investigated in this work.}
\begin{tabular}{ c  c  c  c }
\hline
Area & Long. (\degr) & Lat. (\degr) & Motivation \\
\hline \hline
1 & 20 -- 40 & $-10\rightarrow -4~\&~4\rightarrow 10$ & Synchrotron dominated \\ 
2 & 20 -- 40 & $-$4 $\rightarrow$ 4 & The RRL data region\\
3 & 21 -- 26 & $-$4 $\rightarrow$ 4 & Typical diffuse region  \\
4 & 30 -- 39 & $-$2 $\rightarrow$ 2 & AME region \\
\hline
\end{tabular}
\label{tab:areas}
\end{table}

\subsection{Ancillary Data}

We have compared the C-BASS maps with the all-sky maps at other
frequencies listed in Table~\ref{tab:info}. We will now briefly discuss each data set in turn. All maps 
were used in {\tt{HEALPix}}
format. Those maps that were originally in a different format 
were converted to {\tt{HEALPix}} using a nearest-neighbour pixel 
interpolation. The maps were first smoothed to a common
\mbox{1\degr}\ resolution,
assuming that the original survey beams were Gaussian with the
FWHM specified
in Table~\ref{tab:info}. Depending on the application, the maps were resampled as
required to {\tt{HEALPix}} $N_{\rm side} = 256$ (13\farcm 7 pixels) or 
$N_{\rm side} = 64$ ($55'$ pixels).

\begin{table*}
\setlength{\tabcolsep}{1.5mm}
\caption{The data used in this paper alongside their frequency, FWHM, {\tt{HEALPix}} 
$N_{\rm{side}}$, calibration uncertainty, average thermal noise (in mK Rayleigh-Jeans) unless stated otherwise) and reference. The two bottom rows are derived component separation products.}
\begin{tabular}{ c  r@{}l  r@{}l  c r@{}l  r@{}l  c}
\hline
Survey & $\nu$ & (GHz)& Res.& (arcmin) & $N_{\bf side}$ & Calibration ($\%$)& &$\sigma$&  & Reference \\
\hline \hline
Haslam &  0 & .408 & 51& & 512 & 10&  & 700& & {\protect\cite{Haslam}}\\ 
RRLs & 1 &.4 & 14&.8 & 512 & 15 &  & 5& & {\protect\cite{Alves}}\\
Reich & 1&.420  & 35& & 512 & 10&  & 140& & {\protect\cite{Reich}}  \\
Jonas & 2&.3 & 20&.0 & 256 & 5& & 30& & {\protect\cite{Jonas}}\\
C-BASS & 4&.76 & 43&.8 & 256 & 5&  & 6& & {\protect\cite{King10}}\\
\emph{WMAP}  & 22&.8 & 49& & 512 & 0&.2  & 0&.04 & {\protect\cite{Wmap09}}\\ 
\emph{WMAP}  & 33&.0 & 40& & 512 & 0&.2 & 0&.04 & {\protect\cite{Wmap09}}\\ 
\emph{WMAP}  & 40&.7 & 31& & 512 & 0&.2  & 0&.04 & {\protect\cite{Wmap09}}\\ 
\emph{WMAP}  & 60&.7 & 21& & 512 & 0&.2  & 0&.04 & {\protect\cite{Wmap09}}\\ 
\emph{WMAP}  & 93&.5 & 13& & 512 & 0&.2  & 0&.03 & {\protect\cite{Wmap09}}\\ 
\emph{Planck} & 28&.4 & 32&.65 & 1024 & 0&.4  & 0&.009 &  {\protect\cite{Planck13}}\\ 
\emph{Planck} & 44&.1 & 27&.92 & 1024 &0&.4 & 0&.009 & {\protect\cite{Planck13}}\\ 
\emph{Planck} & 70&.4 & 13&.01 & 1024 &0&.4  & 0&.008& {\protect\cite{Planck13}}\\ 
\emph{Planck} & 143& & 7&.04 & 2048 &0&.4 & 0&.0006 & {\protect\cite{Planck13}}\\ 
\emph{Planck} & 353& & 4&.43 & 2048 & 0&.4 &0&.0003 & {\protect\cite{Planck13}}\\ 
\emph{Planck} & 545& & 3&.80 & 2048 & 7& & 0&.0001 & {\protect\cite{Planck13}}\\ 
\emph{Planck} & 857& &  3&.67 & 2048 & 7&  &0&.00006 & {\protect\cite{Planck13}}\\ 
\emph{COBE}-DIRBE & 1249& & 37&.1 & 1024  & 13&.5 & 0&.5\;MJy\,sr$^{-1}$ & {\protect\cite{COBE}} \\
\emph{COBE}-DIRBE &  2141& & 38&.0 & 1024 & 10&.6 & 32&.8\;MJy\,sr$^{-1}$ &{\protect\cite{COBE}}\\
\emph{COBE}-DIRBE & 2997& & 38&.6 & 1024 & 11&.6 & 10&.7\;MJy\,sr$^{-1}$ &{\protect\cite{COBE}}\\
\hline
& & & & &  Derived\\
\hline
MEM & 22&.8 & 60& & 128 & -& & -& & {\protect\cite{Wmap09}}\\ 
MCMC & 22&.8 & 60& & 64 & -& & - & &{\protect\cite{Wmap09}}\\ 
\hline
\end{tabular}
\label{tab:info}
\end{table*}

\subsubsection{Low frequency data}

We use the \cite{Haslam} 408\,MHz, \cite{Reich}
  1.42\,GHz, and \cite{Jonas} 2.326\,GHz radio maps to provide
  additional low frequency information on synchrotron/free-free
  emission. The Haslam data are used in their unfiltered form so
point sources and striations are still present. These data, alongside
the {\protect\citet{Jonas}} data are also used to determine the
spectral index between 0.408 and 4.76\;GHz outside of the Galactic
plane. The {\protect\cite{Jonas}} data only cover the southern sky
($-83\degr < \delta < 13\degr $) but fortunately the survey extends to
the Galactic plane region under investigation in this work.

\protect\citet{Haslam} estimate an uncertainty of 3\;K in their global
zero level. Although the true random noise is far less than this,
striations present in the data are typically 0.5\;K and worsen to 3\;K
in some regions, so we take 3\;K as a conservative {\em local}
uncertainty in the pixel values.

The temperature scale of the 1.420\,GHz map is problematic, as extensively
discussed by \protect\citet{ReichSpec}. The map is nominally on the full-beam
scale. Because of the near-sidelobe response within 3\fdg 5, fluxes of 
point sources obtained by fitting a Gaussian to the central lobe of the beam 
are expected to be 
underestimates, and this may be corrected by converting to a `main-beam'
brightness temperature. Based on the observed beam profile, this correction
was expected to be about $1.25 \pm 0.15$; however, 
\protect\citet{ReichSpec} find $1.55\pm 0.08$ from direct observations of calibrators,
so either the measured beam area or the original
temperature scale is in error by around 24 per cent. In our 
application we multiplied the data by the empirically deduced correction factor 
of 1.55. We too find this factor to be consistent with our calibration
observations of Barnard's Loop (Section 3). We assign a 
10 per cent uncertainty, mainly
due to the unquantified variation of the temperature scale with the angular
size of the source. We note that \protect\citet{ReichSpec} estimate a full-to-main beam factor of 
$1.04 \pm 0.05$ for the 0.408\,GHz survey, so it suffers much less from
this `pedestal' effect.

To allow for scanning artefacts in the 1.42\,GHz
maps we use a conservative local uncertainty of 0.5\;K
equal to the original estimate of the zero-level uncertainty.

\subsubsection{High frequency data}

We use the \emph{Planck} and \emph{WMAP} 9-year
  data to help constrain the spectral form of AME. The \emph{WMAP}
K-band data are specifically used to determine the ratio of AME to
total emission between 5 and 23\;GHz. The \emph{WMAP} thermal noise in
each {\tt{HEALPix}} pixel ($\sigma$) was calculated using the relation
\protect\citep{Wmap09} $\sigma = \sigma_{0}/\sqrt{N_{{\rm{obs}}}}$,
where $N_{\rm{obs}}$ is the map hit count and $\sigma_{0}$ is the
thermal noise per sample. Therefore the r.m.s. thermal noise values
listed in Table~\ref{tab:info} for \emph{WMAP} are average values
across the sky. The \emph{Planck} 100\;GHz and 217\;GHz maps were not
used as these data include unwanted and significant contributions from
the CO lines {\protect\citep{PlanckCO}}.

An additional 3 per cent uncertainty was assigned to both the
\emph{WMAP} and \emph{Planck} data, as in
     {\protect\cite{PlanckAME11}}, to account for residual beam
     asymmetries after smoothing.  Both \emph{WMAP} and \emph{Planck}
     have unblocked and severely under-illuminated apertures to reduce
     sidelobes to extremely low levels, and so the difference between
     the antenna temperature and full-beam brightness temperature
     scale is less than 0.5 per cent (in the case of \emph{WMAP}, the
     data are corrected for far sidelobe contamination and the
     corrections from the \emph{WMAP} `main-beam' to the C-BASS
     full-beam are less than 0.6 per cent {\protect\citep{Jar2007}},
     and we have not made any correction in this analysis.

We use the \emph{COBE}-DIRBE band 8, 9 and 10 data to help constrain
the spectral form of thermal dust emission. The \emph{COBE}-DIRBE data
used are the Zodi-Subtracted Mission Average (ZSMA) data
{\protect\citep{COBE}}.

\subsubsection{Free-Free templates}

At low Galactic latitudes ($|b| < 4\degr$), the intensities of the
diffuse free-free and synchrotron emission are comparable at gigahertz
frequencies.  We separate the synchrotron and free-free contributions
to total emission in the Galactic plane at 5\;GHz; to do this a
free-free emission template was required as well as the C-BASS data.
Free-free emission is unpolarized and characterised by a spectral
index of $ \beta \approx - 2.1$ {\protect\citep{Gold2009, CliveFF}},
for optically thin regions. At high frequencies \mbox{($\approx$
  90\;GHz)}, the free-free spectral index will steepen
{\protect\citep{ISM}} to $-2.14$ due to the Gaunt factor, which
accounts for quantum mechanical corrections, which become important at
higher frequencies.

The source of the diffuse Galactic free-free emission is the Warm
  Ionized Medium (WIM), with its intensity proportional to the 
  emission measure (EM), given by $\rm{EM} = \int \left(n_{e}\right)^{2} \; dl$,
where $n_{e}$ is the electron density.  The
WIM also radiates $\rm{H} \alpha$ recombination lines, including
the optical Balmer-alpha line (656.28\;nm) and high-n radio
recombination lines (RRLs) whose strength also depends on the EM.  As a result,
$\rm{H} \alpha$ lines can be used to calculate free-free emission
intensities provided the electron temperature is known. 
However, $\rm{H}\alpha$ lines are not satisfactory free-free tracers in the
Galactic plane: dust absorption along the line-of-sight of the $\rm{H}
\alpha$ lines (which needs to be corrected for throughout the Galaxy)
reaches a maximum in the plane due to the increased density of
dusty star forming regions. RRLs from ionised hydrogen provide
an absorption-free alternative for tracing free-free emission in
the Galactic plane {\protect\citep{Alves}}. We derive the free-free template used in this analysis
from RRL data constructed from the HI Parkes All-Sky Survey
(HIPASS; {\protect\citealt{hipass}}).  

The RRL data have a beam FWHM of 14.8 arcmin, and
have thermal noise and calibration uncertainties of 5\;mK and 15 per
cent respectively. The RRL data {\protect\citep{Alves}} cover $20\degr
< l < 40\degr, -4\degr < b < 4\degr$ and were used by Alves et al. to
produced a free-free emission template from the data by using an
average electron temperature of 6000\;K across the whole region.  This
average value carries a large uncertainty ($\pm$ 1000\;K), as it does
not account for the presence of warmer and cooler regions. This
uncertainty is the dominant factor in the 15 per cent uncertainty
assigned to the free-free template.

For this analysis the RRL data provide our free-free template of
choice.  As this is a novel approach to deriving free-free templates,
we find it illuminating to compare the RRL-derived result with those
from \emph{WMAP}. All-sky free-free models are available from the
\emph{WMAP} \mbox{9-year} product release constructed using a maximum
entropy method (MEM) and Markov Chain Monte Carlo simulation
(MCMC)\footnote{\url{http://lambda.gsfc.nasa.gov/product/map/dr5}}. The
MCMC free-free model \citep{Gold2009} at 22.8\,GHz used in this work
(`model f') fits the total emission model for each
  pixel as a sum of a power-law with a free spectral index parameter
  for synchrotron emission, a fixed-index power-law for the free-free
  emission, a CMB term, a power-law with free spectral index for the
  thermal dust emission, and a theoretical spinning dust curve with
  amplitude. The MEM free-free model at 22.8\,GHz is a similar
  spectral pixel-by-pixel fit, which uses Bayesian priors for pixels
  where the data do not constrain the parameters very well (see
  \citealt{Gold2009} for details).


\section{Data Validation}
\label{sec:dataVal}

We have considered the effect of pointing errors, absolute intensity
calibration, colour (bandpass) corrections and atmospheric opacity on
the C-BASS temperature scale, but the dominant error is the
uncertainty on the aperture efficiency, which we estimate at \mbox{5
  per cent}. To assess the consistency of the C-BASS
  temperature scale, we compare C-BASS data with other radio
  surveys. We focus on Barnard's Loop, an H{\sc{ii}} shell within the
  Orion complex {\protect\citep{BsLoop}}. This region is known to be
  dominated by optically thin free-free emission at these frequencies,
  which allows us to use the well-defined spectral index to compare
  data at a range of frequencies. Fig.~\ref{fig:BsLoop} shows
  Barnard's Loop in H$\alpha$ Balmer line emission and in
  \mbox{4.76\;GHz} radio continuum as seen by C-BASS North.The boxed region shows the specific area selected for analysis in Fig. ~\ref{fig:TTBs}.
  The close
  morphological similarity indicates that free-free emission from warm
  ionized gas dominates.

The temperature spectrum of a diffuse Galactic emission is generally
approximated in the form of a power-law:
\begin{equation}
T(\nu) \propto \; \nu^{\beta},
\end{equation}
where $T(\nu)$ represents the brightness temperature at frequency
$\nu$ and $\beta$ is the spectral index of the temperature
distribution. Given two intensity data sets taken at different
frequencies ($\nu_{1}$ and $\nu_{ 2}$) a temperature-temperature
(\mbox{$T$-$T$}) plot of these two sets will reveal a linear
relationship, the gradient of which relates to the emission spectral
index \mbox{{\protect\citep{TTplot}}} via:
\begin{equation}
T(\nu_{1}) = \left(\frac{\nu_{1}}{\nu_{2}}\right)^{\beta} T(\nu_{2}) \; + \; \rm{baseline \; offsets}.
\end{equation}
An advantage of the $T$-$T$ plot method is its insensitivity to
zero-level uncertainties, though if several emission mechanisms are present several linear relationships will be seen so the region for analysis must be chosen with care. Optically thin free-free emission has a
well-documented spectral index of $-2.12 \pm 0.02$
{\protect\citep{ISM}}, therefore a simple and effective validation of
the \mbox{C-BASS} data quality can be achieved using $T$-$T$ plots of
this area.

The compact sources Orion A (M42), Orion B and NVSS J060746-062303
have been masked out of the C-BASS image using a diameter $3.5 \times$
the C-BASS FWHM for Orion A and B and $2 \times$ the C-BASS FWHM for
NVSS J060746-062303 (see Fig.~\ref{fig:BsLoop}). A
smaller mask size could be used for NVSS
J060746-062303 as the source is fainter than Orion A or Orion B.

\begin{figure}
\centering 
\includegraphics[width=0.4\textwidth]{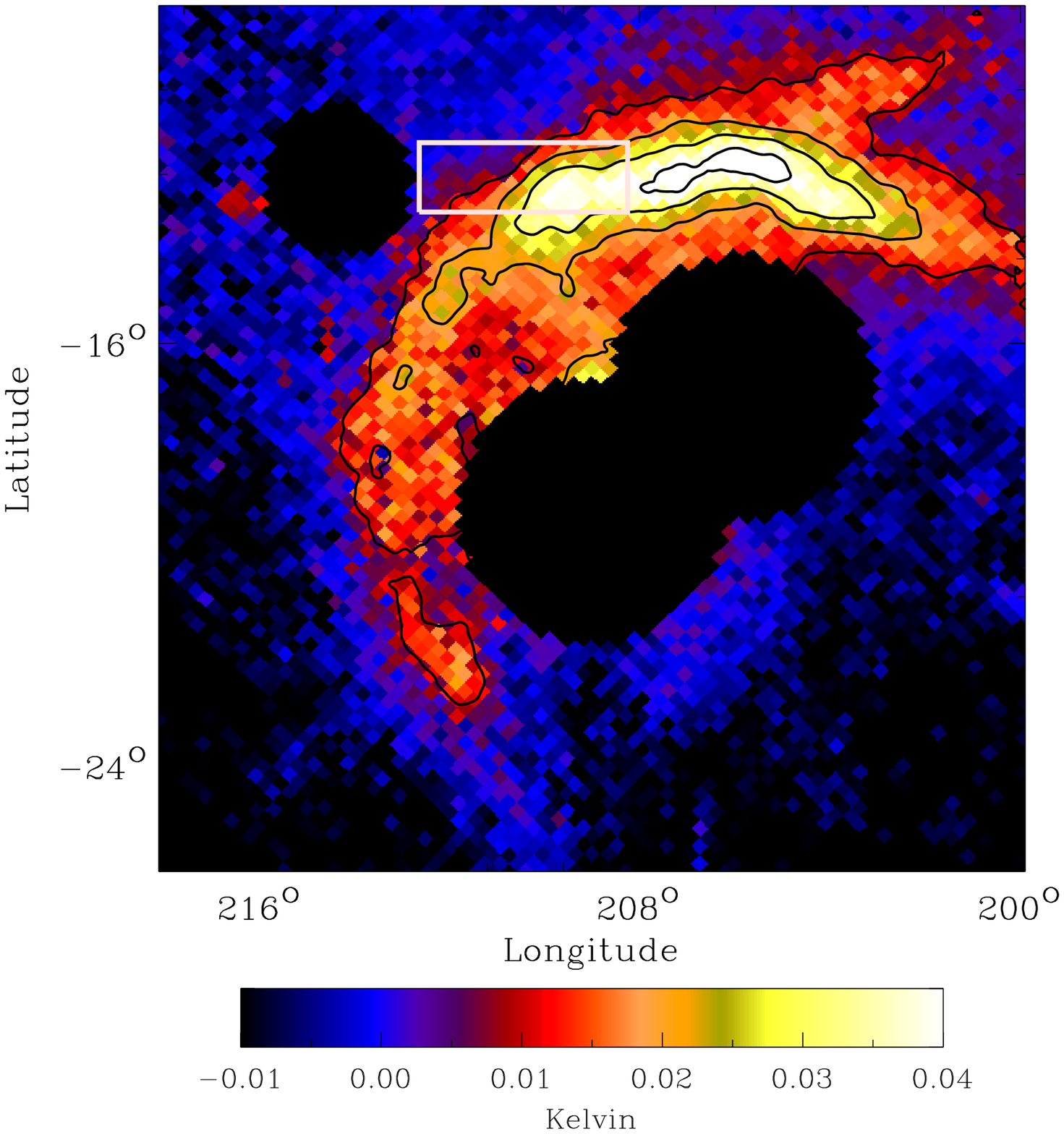} 
\includegraphics[width=0.4\textwidth]{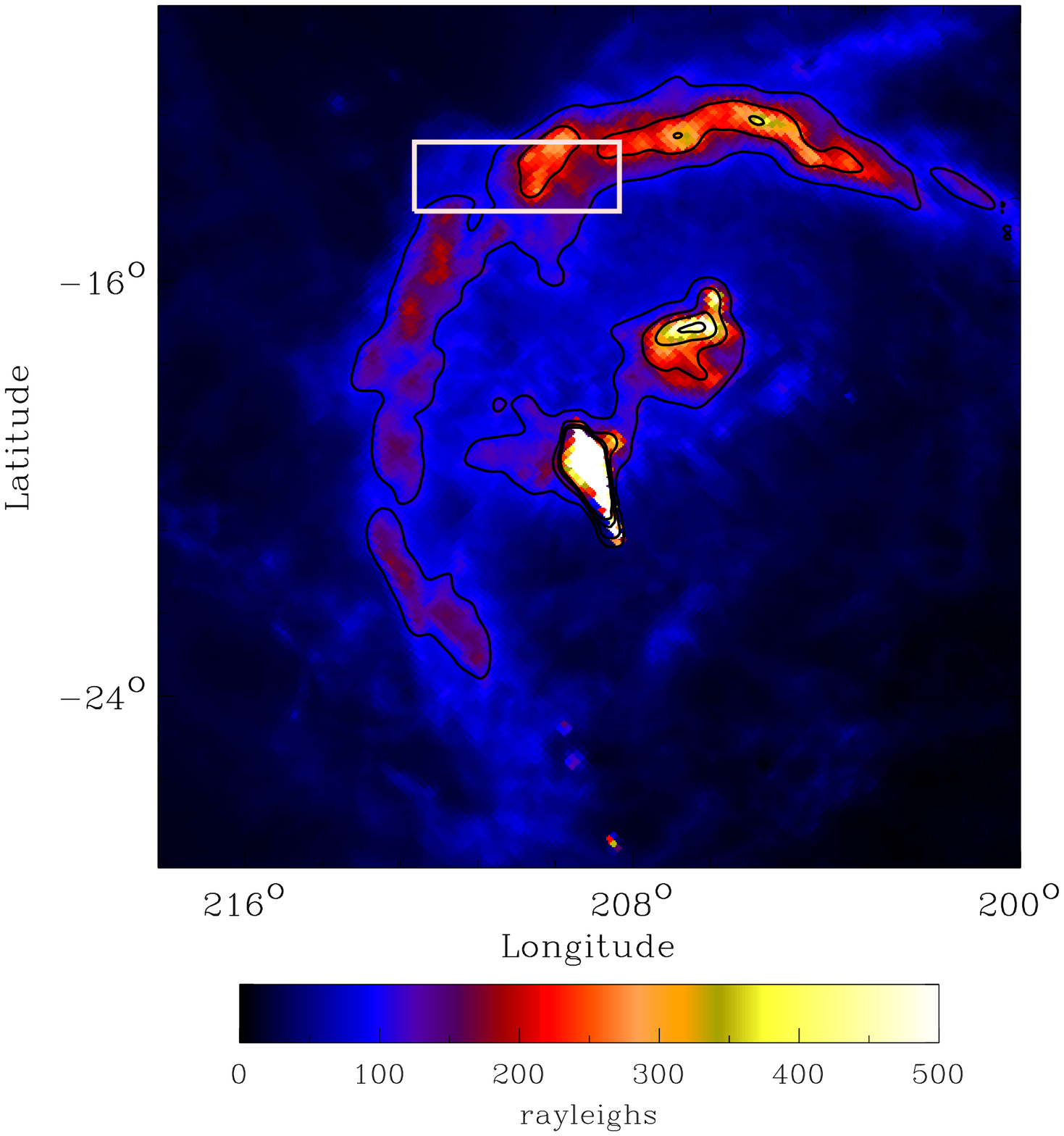}
\caption{{\it Top}: Barnard's Loop as seen at 4.76\;GHz by C-BASS
  North. $N_{\rm side} = 256$ and FWHM resolution of 0\fdg 73. Orion
  A, Orion B and NVSS J060746-062303 are masked out. The colour scale
  in in kelvin and the contours are spaced at 0.01\;K
  intervals. {\it{Bottom}:} Barnard's Loop in H$\alpha$ emission at
  {\tt{HEALPix}} $N_{\rm side} = 512$ and resolution 6.1 arcmin. The
  colour scale is in rayleighs and the contours spaced at 100 rayleigh
  intervals between 100 and 500
  {\protect\citep{Halpha}}. The morphological
    similarity between the maps suggests that free-free emission is
    dominant at 4.76\,GHz. The white box encapsulates the region selected for temperature analysis.}
\label{fig:BsLoop}
\end{figure}

Fig.~\ref{fig:TTBs} displays $T$-$T$ plots using the C-BASS data for
Barnard's Loop against {\protect\citet{Haslam}},
{\protect\citet{Reich}} and \emph{WMAP} K-band data. The $T$-$T$ plots
are made at $N_{\rm side} = 64$ to reduce correlations
between the pixels -- at a FWHM of 1\degr, this gives $< 2$ pixels per
beam.

The error bars in Fig.~\ref{fig:TTBs} represent the pixel noise for the respective radio maps calculated using 1000 iteration Monte Carlo
simulations. These include the effects of confusion noise (0.8\;mK for the
\mbox{C-BASS} data), zero-level uncertainties (3\;K for the {\protect\citet{Haslam}} and 0.5\;K for the
{\protect\citet{Reich}} data), CMB anisotropies (for the \emph{WMAP} data),
{\tt{HEALPix}} smoothing and degrading as well as the r.m.s. thermal noise. The CMB anisotropies
were simulated using the {\tt{CAMB}} web interface\footnote{\url{http://lambda.gsfc.nasa.gov/toolbox}} and the standard, pre-set cosmological parameters. 
It was determined that calibration uncertainty, and not pixel noise, was the
dominant source of uncertainty. The uncertainties quoted for the spectral indices are a
quadrature combination of the pixel errors and calibration
uncertainties associated with each data set. 

The 0.408--4.76\;GHz, 1.42--4.76\;GHz and 4.76--22.8\;GHz $T$-$T$ plots
in Fig.~\ref{fig:TTBs} show linear fits; the reduced $\chi^{2}$ values
are 1.9, 0.5 and 1.3 for the 0.408--4.76\;GHz, 1.42--4.76\;GHz and
4.76--22.8\;GHz data, respectively. The average spectral index 
of $\beta = -2.15 \pm 0.03 $ confirms a free-free dominated spectrum 
for Barnard's Loop. This spectral index is
consistent within 1$\sigma$ of the expected value ($\beta \approx -2.12$). 

The multiplicative factor required to bring C-BASS data to be in perfect agreement with \emph{WMAP} 22.8\,GHz data (assuming all the emission is due to free-free emission) is $0.95\pm0.05$. This validation gives us
confidence that there are no significant systematic calibration
errors in excess of the 5 per cent calibration error assigned to the
preliminary C-BASS data.

\begin{figure}
\includegraphics[width=0.48\textwidth]{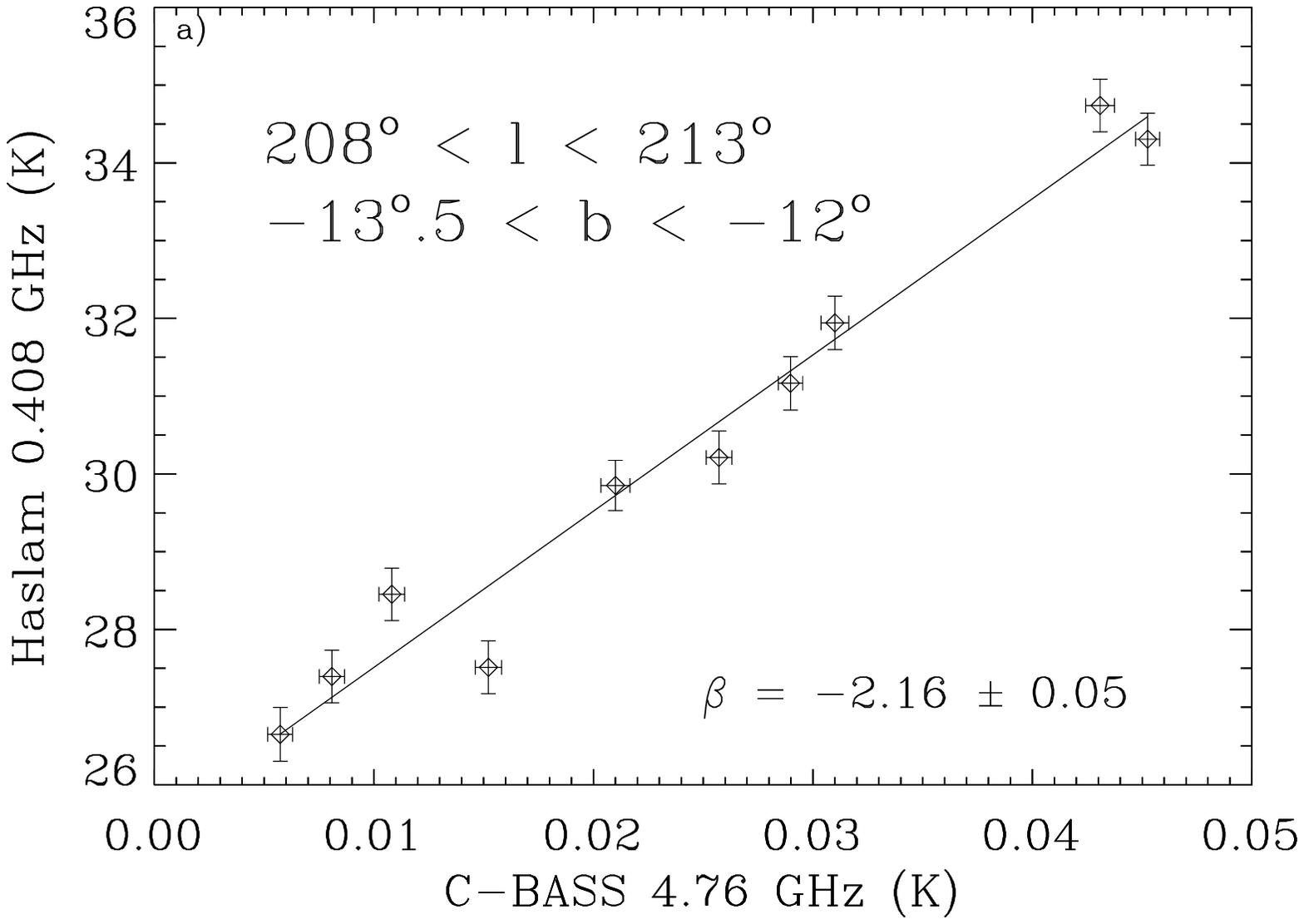} 
\includegraphics[width=0.48\textwidth]{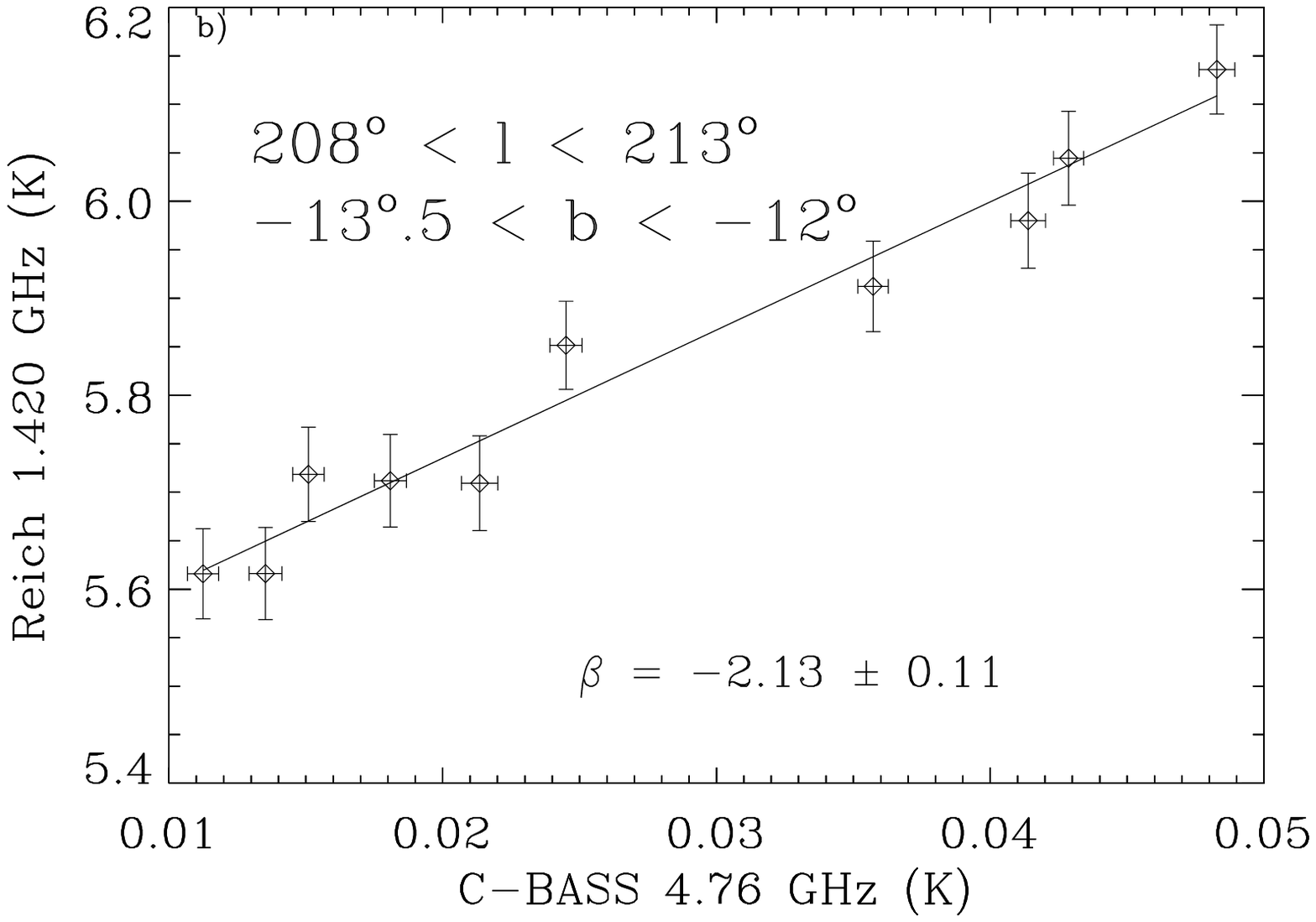}
\includegraphics[width=0.48\textwidth]{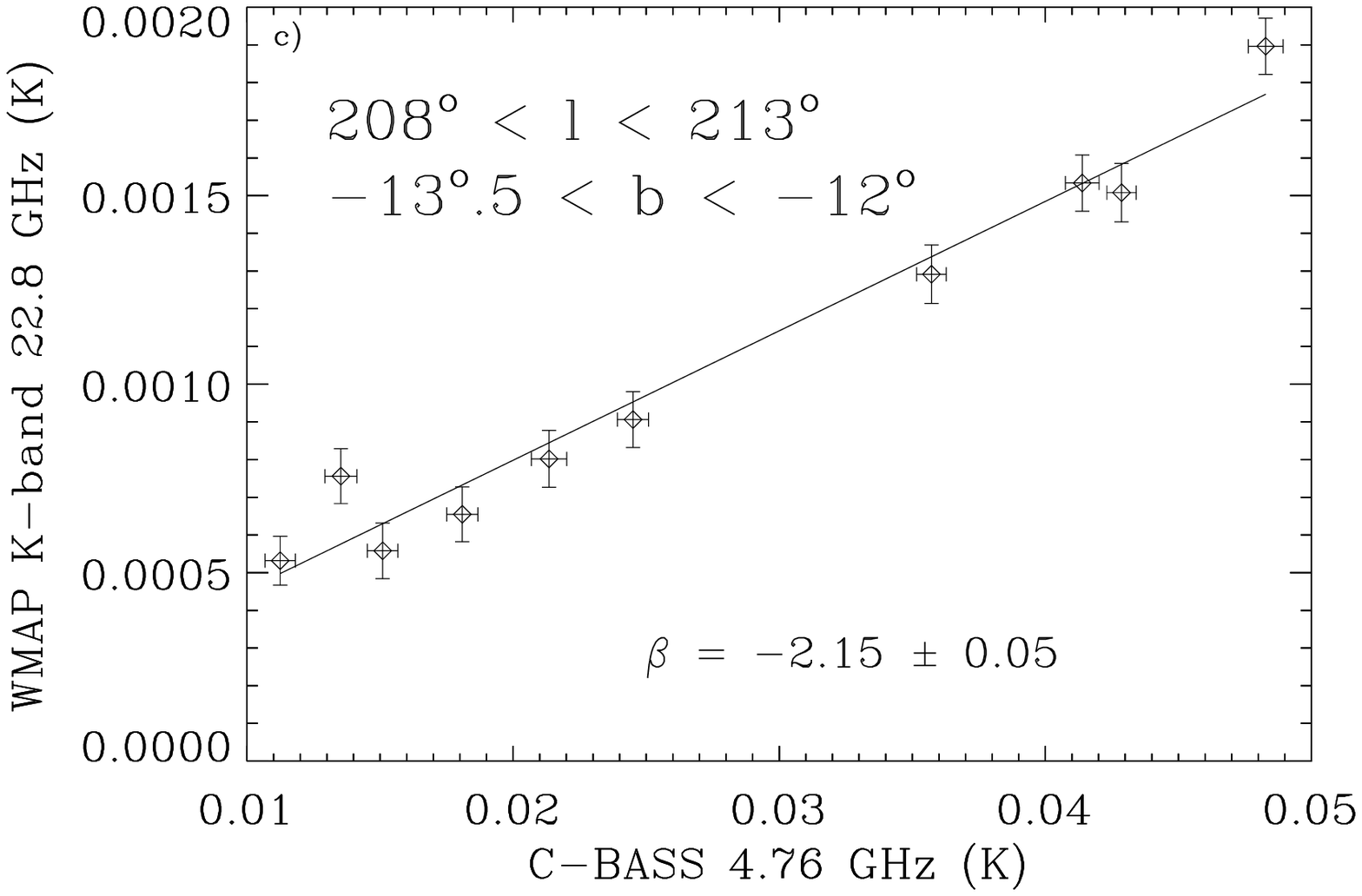}
 \caption{$T$-$T$ plots of the Barnard's Loop region between 4.76\;GHz and (a) 0.408\,GHz, (b) 1.42\,GHz and (c) 22.8\,GHz. The spectral indices confirm a free-free emission dominated region.} 
 \label{fig:TTBs}
\end{figure}

\section{Diffuse emission between 0.408 and 4.76\,GH\lowercase{z}}
\label{sec:synchrotron}

The total diffuse Galactic plane radio continuum 
emission seen at 4.76\;GHz is primarily a
mixture of free-free and synchrotron emission. Free-free emission 
is known to have a narrow latitude
distribution while synchrotron emission is broader {\protect\citep{Alves,RodsSync}}. The narrow width of the
free-free distribution shown in Section ~\ref{sec:freeLon}
suggests that the free-free contribution is negligible compared
to the synchrotron contribution at latitudes higher than four degrees. 
This suggestion is supported by
the  H$\alpha$ free-free template {\protect \citep{CliveFF}},
 which covers off-plane regions. It is therefore possible 
to determine the spectral index of synchrotron emission 
between 0.408 and 4.76\;GHz at intermediate and high Galactic
latitudes without having to perform any free-free/synchrotron
 emission separation.

\subsection{Intermediate latitude total emission spectral indices}

The spectral index of the Galactic synchrotron emission varies
 significantly, both spatially and with frequency. 
{\protect\citet{Strong}} review results from sky maps 
between 22\;MHz and 23\;GHz at intermediate latitudes, finding  a steepening 
in the synchrotron spectrum from $\beta\approx -2.5$ at 0.1\,GHz to $\beta \approx -3$
at 5\,GHz.  
Between 0.408\,GHz and 3.8\,GHz an index of $\approx -2.7$ has been
estimated
in the Galactic plane whereas between 1.42 and 7.5 GHz the
spectrum appears to steepen to $\approx -$3.0
{\protect\citep{Platania98}}. {\protect\citet{Jaffe}}, {\protect\cite{Lawson}} and
{\protect\cite{Bennett} also discuss the spectral steepening of synchrotron
emission in the Galactic plane and it is clear that between 2.3 and 22.8\;GHz the spectral index
undergoes a steepening from $\approx -2.7$ to $\approx -3.1$. Due to the
relationship between electron energy distribution $N(E) \propto E^{\;\delta}$, and synchrotron spectral index  $\beta = -(\delta + 3)/2$, this spectral index steepening
 results from a steepening of the cosmic ray electron spectrum.
It resembles that expected from synchrotron losses, i.e.  a `break' (actually
a rather smooth steepening) of 0.5--1 in $\beta$ assuming a steady state 
between injection of fresh particles, radiative loss, and transport out
of the Galaxy \protect\citep[e.g.][]{BulanovDogel,Strong}. However, the timescale for electrons to leave
the Galaxy implied by this interpretation is far shorter than the 
cosmic ray residence timescales inferred from the presence of spallation 
product nuclei in the cosmic rays, so in current models 
\protect\citep[e.g.][]{OrlandoStrong} the steepening is
explained by a break in the electron injection energy spectrum;
the break due to radiative losses is predicted to occur at a much 
lower frequency than is actually observed.

The steepening of the synchrotron spectrum at a few GHz is consistent
with the measured steepening of the local cosmic ray electron spectrum 
\protect\citep{fermilat}, with $\delta$
varying from $<2.3$ to $\approx 3$ between 1 and 10 GeV 
(solar modulation prevents accurate measurement of $\delta$ much below 1 GeV).

\protect\citet{ReichSpec} observed a steepening along the
Galactic plane of the synchrotron $\beta$ (between 
0.408 and 1.420\,GHz)  with $\beta = - 2.86$ at $ (l,b) = (20\degr,
0\degr)$ and \mbox{$\beta = -2.48$} at $ (l,b) = (85\degr,
0\degr)$.
\protect\citet{PeelSync} comment on the lack of flatter-spectrum 
($\beta \approx -2.7$) emission at high latitudes
between 2.3\;GHz and 30\;GHz, in contrast to the Galactic 
plane.  Conversely, the most sophisticated models of cosmic
ray propagation in the Galaxy to date \protect\citep{OrlandoStrong} 
predict variations in $\beta$ of no more than 0.1 between different
directions, e.g. between the plane and higher latitudes. These models
are explicitly of the large-scale diffuse emission only, and so will omit
any spectral variations caused by features on sub-kpc scales, including
the radio ``loops'' that dominate the high-latitude synchrotron emission. These
are expected to cause variations in $\beta$ because they may be sites
of particle injection, hence with flatter spectra (as for most supernova
remnants) and because they have stronger magnetic fields, so emission
at a given frequency is from lower energy electrons (which will also
flatten the spectrum). Spectra {\em steeper} than the diffuse emission
require that they contain old electrons confined by a very low effective
diffusion coefficient in magnetic fields much stronger than typical interstellar fields, 
so that  significant {\em in situ} radiative losses can occur without compensating
particle acceleration.  It is also worth noting
that the low-latitude regions claimed to show flatter synchrotron spectra
are the ones most contaminated by free-free emission, e.g. the Cygnus X
region at $l =75\degr$--$86\degr$. 

Fig.~\ref{fig:SyncTT} shows $T$-$T$ plots of semi-correlated pixels (6 pixels per beam), which were made in the
off-plane latitude range of $4\degr < |b| < 10\degr$ with no masking of specific regions. We expect this region to be dominated by synchrotron emission
between 0.408/1.420\,GHz and 4.76\,GHz. The positive and negative
latitude ranges are plotted separately to differentiate between
Galactic plane emission and emission from the North Polar Spur and the
Gould Belt. The region in Figs.~\ref{fig:SyncTT}b and \ref{fig:SyncTT}d encompass part of the Gould Belt and the North
Polar Spur. The variation in synchrotron spectral index across
the positive latitude range results in a bulged point distribution in Fig.~\ref{fig:SyncTT}b
between 0.12 and 0.16\;K on the x-axis. 

\begin{figure*}
 \includegraphics[width=0.48\textwidth]{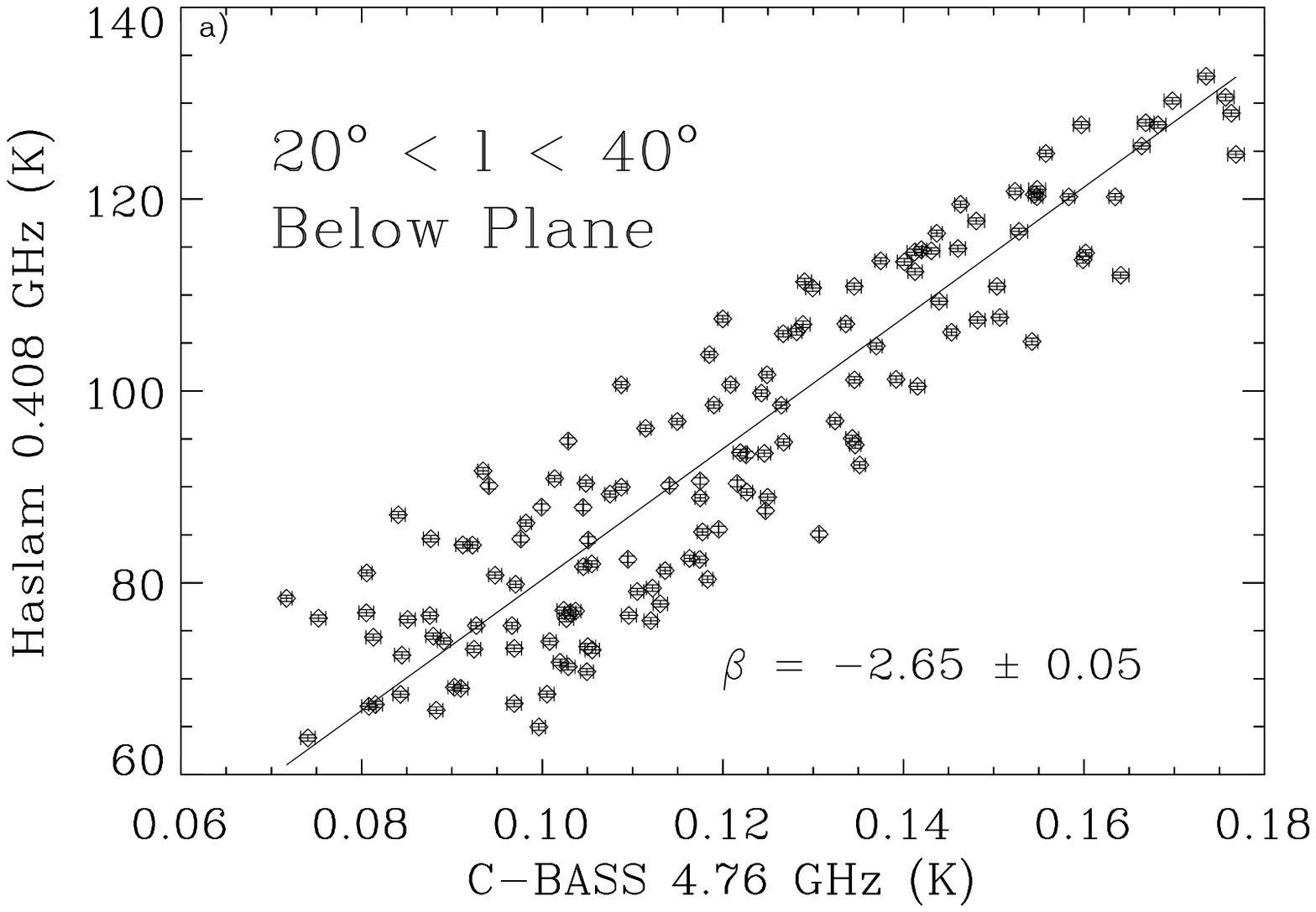} 
 \includegraphics[width=0.48\textwidth]{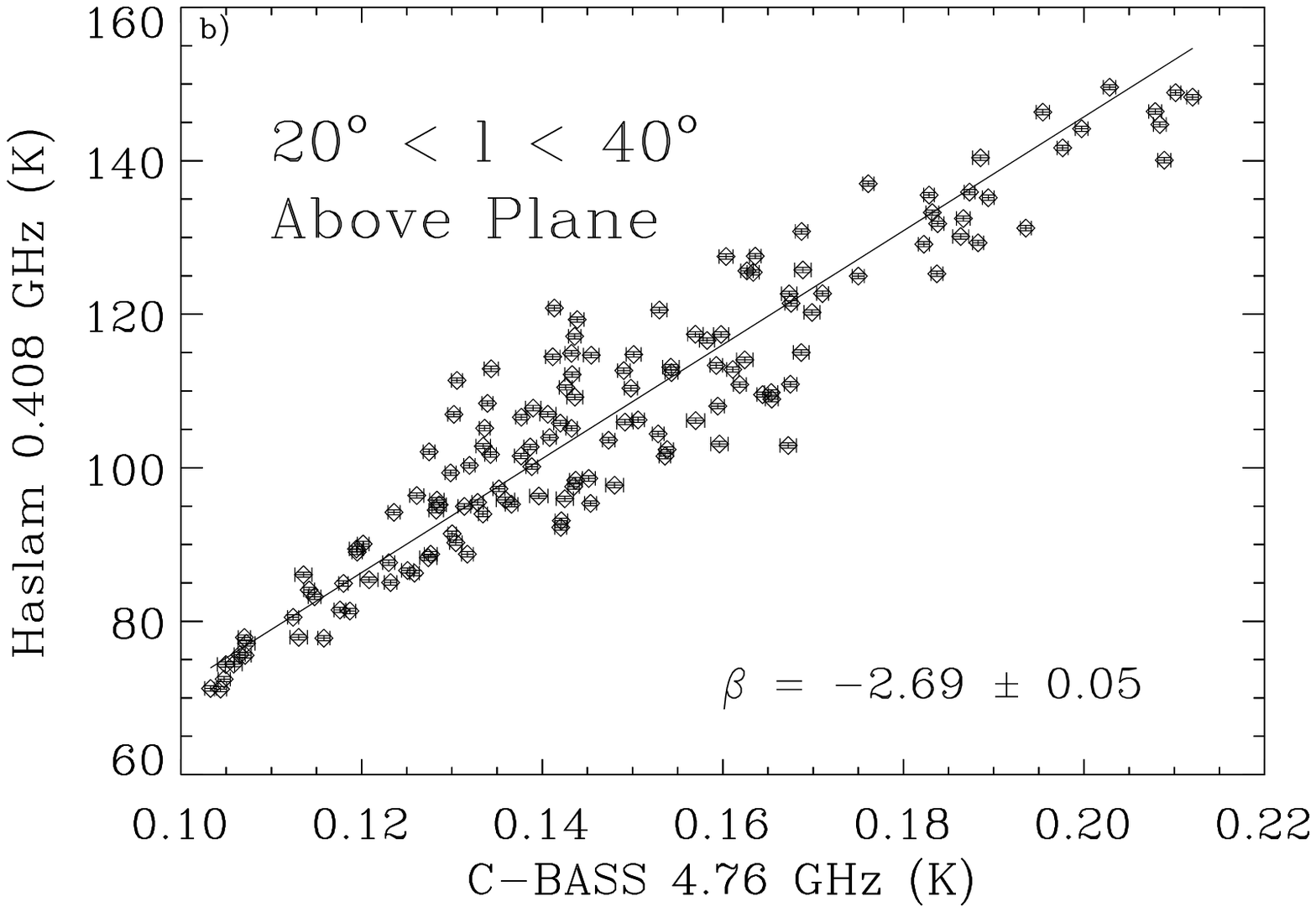}
 \includegraphics[width=0.48\textwidth]{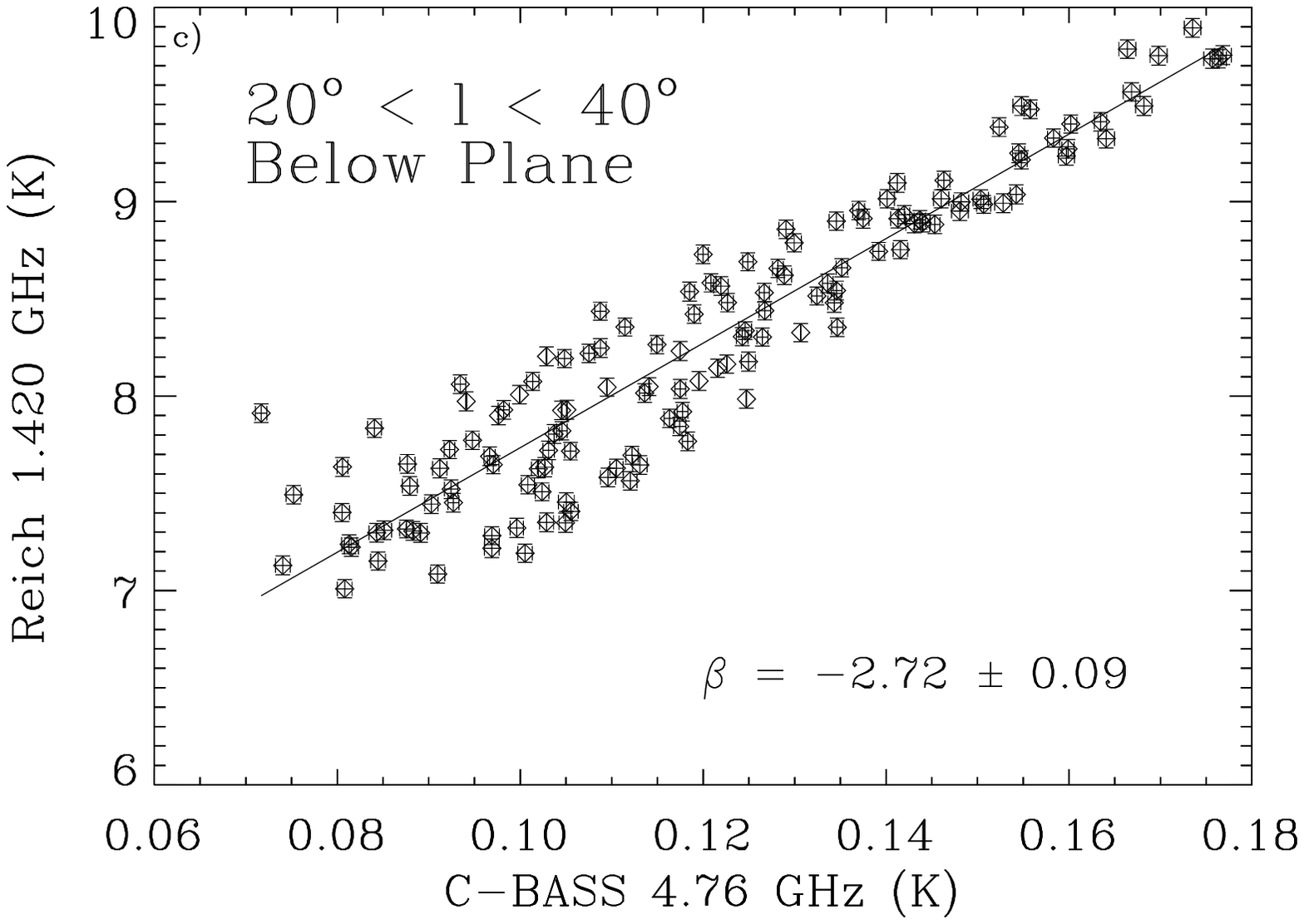} 
 \includegraphics[width=0.48\textwidth]{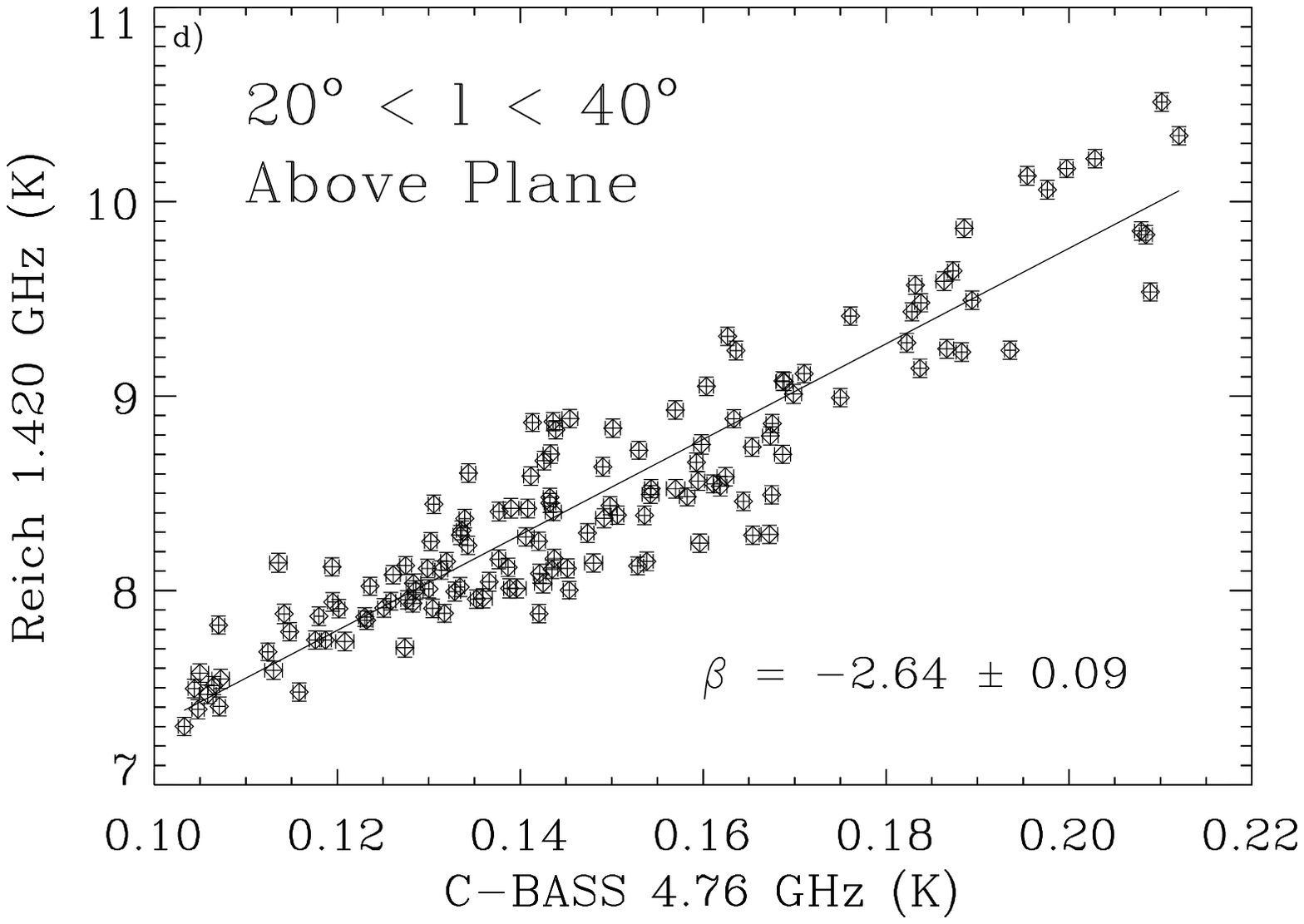}
  \caption{{\it{Top row:}} $T$-$T$ plots between 0.408\;GHz and 4.76\;GHz for (a) $-10\degr < b < -4\degr$ (Below Plane) and (b) $10\degr > b > 4\degr$ (Above Plane). {\it{Bottom row:}} $T$-$T$ plot for 1.420\,GHz and 4.76\,GHz for (c) $-10\degr < b < -4\degr$ and (d) $10\degr > b > 4\degr$. The off-plane spectral indices between 0.408 and 4.76\;GHz can be seen to range between  $-2.67$ and $-2.72$.} 
  \label{fig:SyncTT}
\end{figure*}

The mean spectral indices shown in Fig.~\ref{fig:SyncTT} are
summarised in Table~\ref{tab:specValTab} and range between $ \beta =
-2.64$ and $\beta = -2.72$. The weighted mean spectral indices between
0.408--4.76\,GHz and 1.420--4.76\;GHz are $-2.67\pm0.04$ and
$-2.68\pm0.06$, respectively. These results are
  consistent with, but more accurate than, those in the literature,
  which typically give an average synchrotron spectral index of
  $\approx -2.7$ both within the galactic plane and away from the
  radio ``loops''. The C-BASS data provide a longer
  lever-arm for constraining spectral indices with uncertainties
  $\Delta \beta < 0.1$. The larger uncertainties on the 1.420--4.76
spectral indices are due to the larger uncertainties
  in the 1.42\,GHz survey. We expect the full C-BASS survey to have a
significantly reduced uncertainty in its calibration, and in
conjunction with other modern surveys such as S-PASS
\protect\citep{spass} we should be able to constrain the spectral
index and possible curvature (steepening/flattening) of the spectrum
even more accurately.

\begin{table}
\centering
\caption{The off-plane ($4\degr < |b| < 10\degr$) spectral indices between 0.408/1.420\,GHz and 4.76\,GHz.}
 \begin{tabular}{ c  c  c  c  c }
\hline
Lon. ($\degr)$ & Lat. (\degr) & $\nu$ (GHz) & $\beta$ \\
\hline \hline
20 -- 40 & $-10$ $\rightarrow -4$ & 0.408 -- 4.76 & $-2.65 \pm 0.05$ \\ 
20 -- 40 & 4 $\rightarrow$ 10 & 0.408 -- 4.76 & $-2.69 \pm 0.05$ \\
20 -- 40 & $-10 \rightarrow -4$ & 1.420 -- 4.76  & $-2.72 \pm 0.09$\\
20 -- 40 & 4 $\rightarrow$ 10 & 1.420 -- 4.76 & $- 2.64\pm 0.09$ \\
\hline
\end{tabular}
\label{tab:specValTab}
\end{table}

\subsection{Free-free cleaned maps at 0.408/1.420/4.76 GHz }
\label{sec:freeLon}

At low Galactic latitudes ($|b| < 4\degr$) the free-free emission
 cannot be assumed to be  negligible; the synchrotron and
free-free components need to be separated. Assuming that below
\mbox{5\;GHz} only synchrotron and free-free emission are present at a
detectable level, synchrotron maps can be formed from the
{\protect\citet{Haslam}}, {\protect\cite{Reich}} and C-BASS data using a pixel-by-pixel 
subtraction of a free-free template of choice. More sophisticated component 
separation methods, such as parametric fitting, could be used to acquire synchrotron maps 
but this would require an accurate estimate of the C-BASS zero-level. Therefore, until the
full survey is complete we use simpler methods of component separation.  

 \begin{figure}
\centering 
\includegraphics[width=0.48\textwidth]{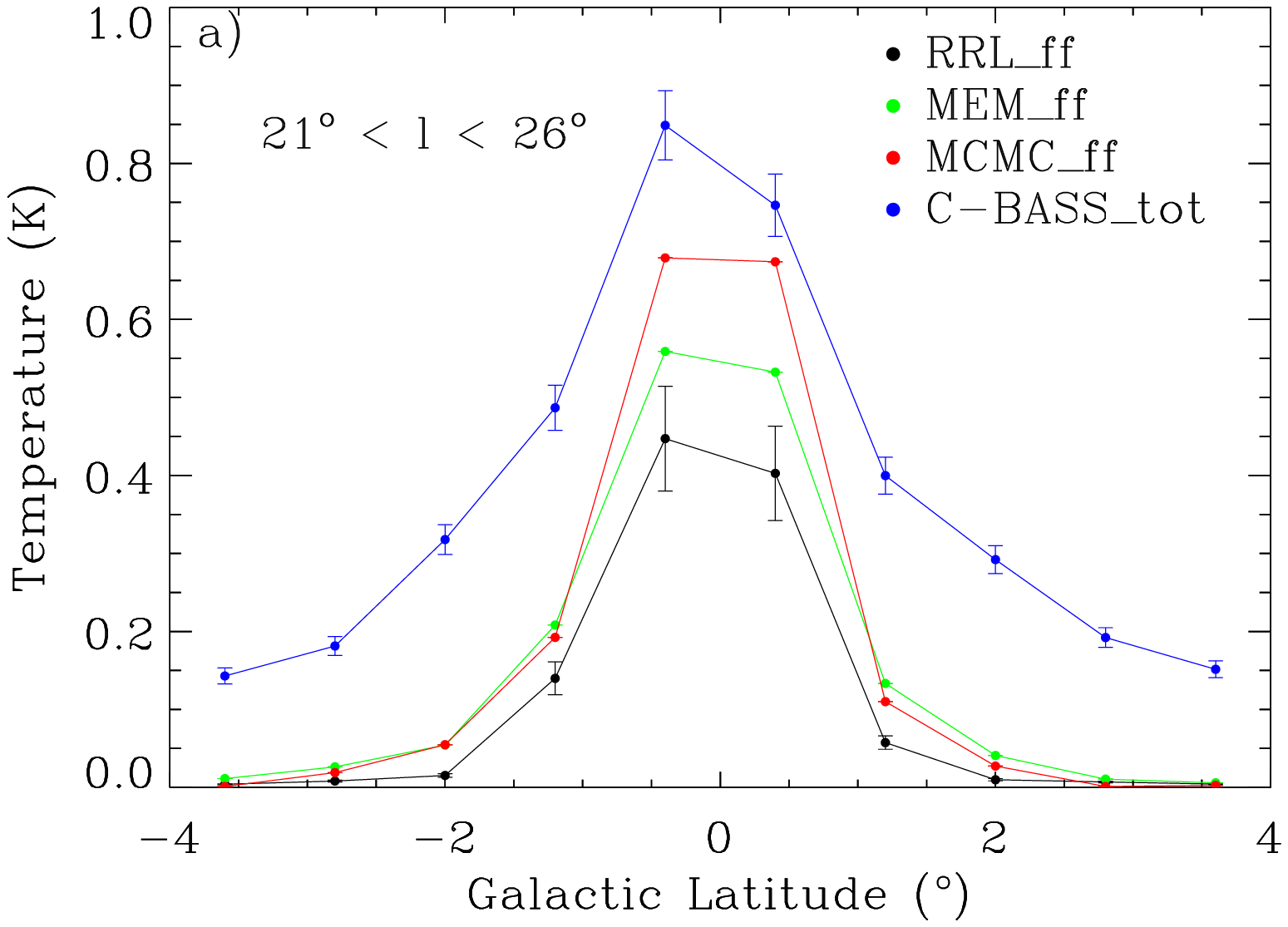} 
\includegraphics[width=0.48\textwidth]{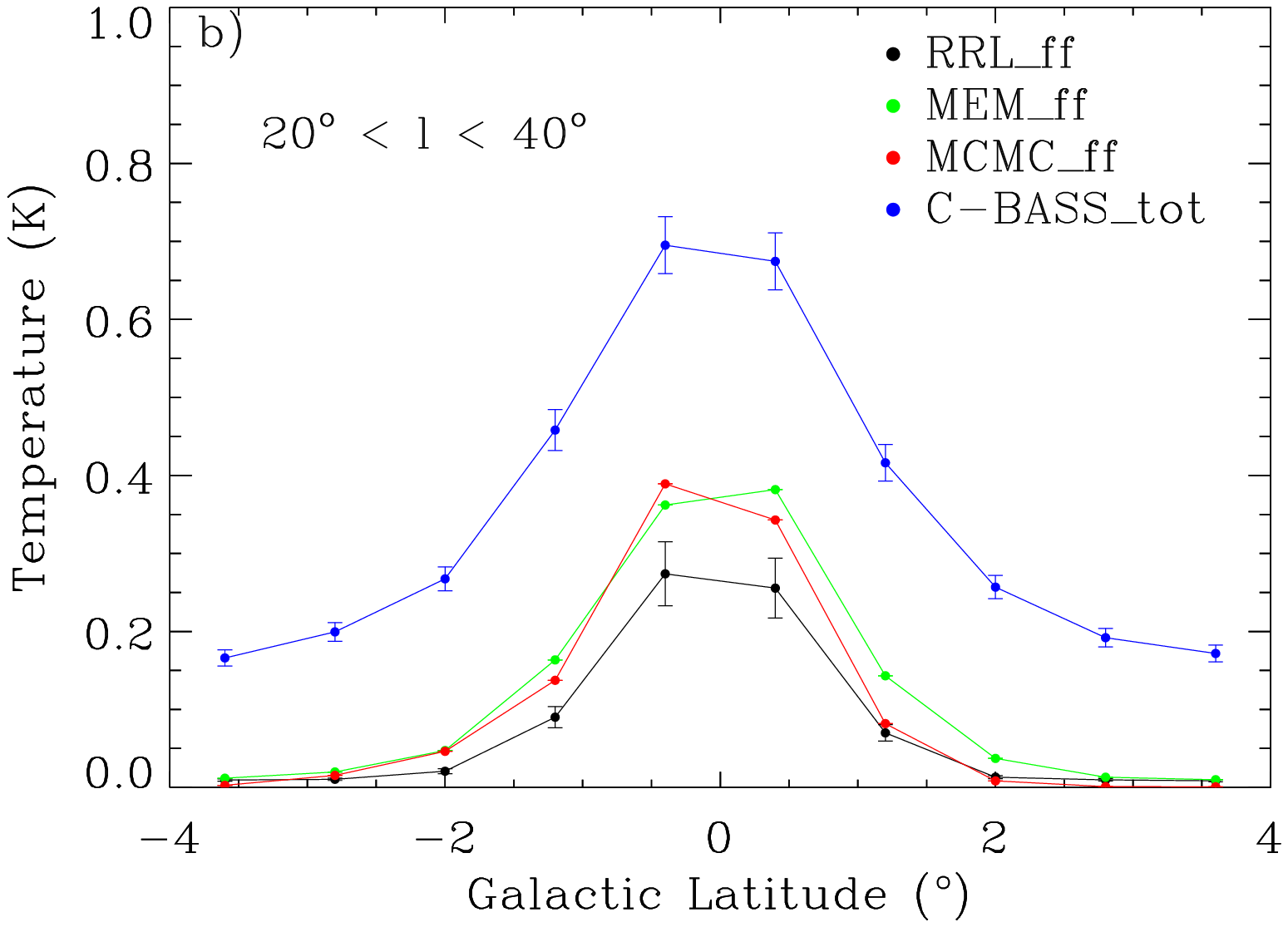}
\caption{Free-free latitude distribution for $|b| < 4\degr$ using averaged longitude data from the RRL, MCMC, MEM free-free templates all scaled to 4.76 GHz for the longitude range of a) 21\degr\ -- 26\degr\ and b) 20\degr -- 40\degr. The maps have been smoothed to 1\degr\ resolution and downgraded to {\tt{HEALPix}} $N_{\rm side} = 64$.}
  \label{fig:free}
\end{figure}

In Fig.~\ref{fig:free}, we present two latitude
  profiles for the RRL, MEM, and MCMC free-free templates over-plotted
  on the \mbox{C-BASS} data for the same region.  The RRL, MCMC and MEM
  templates have been scaled to 4.76\,GHz using the free-free spectral index of
  \mbox{$\beta$ = $-$2.1}. The
  free-free profiles all have their baselines subtracted so that the
  in-plane emissions can be compared without biasing from different
  survey zero-levels. The baseline level was established using a
  cosecant ($1/{\rm sin}(|b|)$) plus a slope to the $|b| < 10\degr$ points for each of the curves. We then subtract the slope (offset and gradient) to leave the plane emission. The 1\degr\ FWHM maps were resampled at {\tt{HEALPix}} \mbox{$N_{\rm side} = 64$}
to match the MCMC template. The two latitude profiles average over a) $21\degr < l < 26\degr$ and
b) the full RRL longitude range $20\degr < l < 40\degr$. The 21\degr\ to 26\degr\ longitude range was selected to be a
 typical bright diffuse region as it contained the fewest bright, compact areas in the C-BASS map.
Both profiles span the full RRL latitude range of
$-4\degr < b < 4\degr$.

The C-BASS data were used as a `testbed' for these templates
and help to identify regions over which they fail to describe the
measured emission. The RRL free-free template is systematically
lower in temperature than the MEM and MCMC templates because the electron 
temperature used is 1000\;K lower than the 7000\;K assumed by 
\emph{WMAP}. The RRLs are a more direct free-free measure
as they are not subject to the same parameter degeneracies faced 
by the MCMC and MEM templates but of course are so far only available 
for $|b| < 4\degr$ and a limited range of longitudes.

 \begin{figure}
\includegraphics[width=0.48\textwidth]{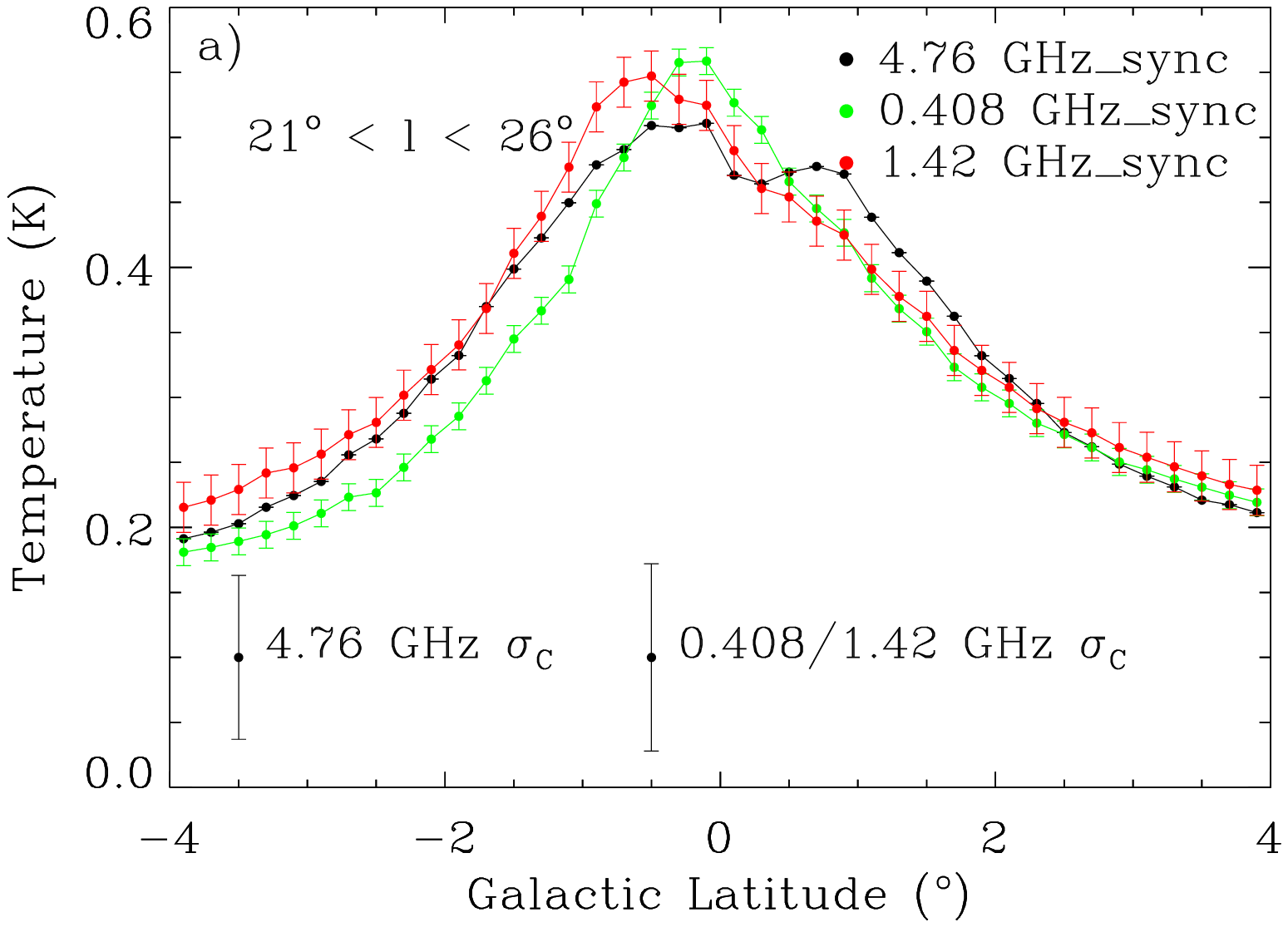} 
 \includegraphics[width=0.48\textwidth]{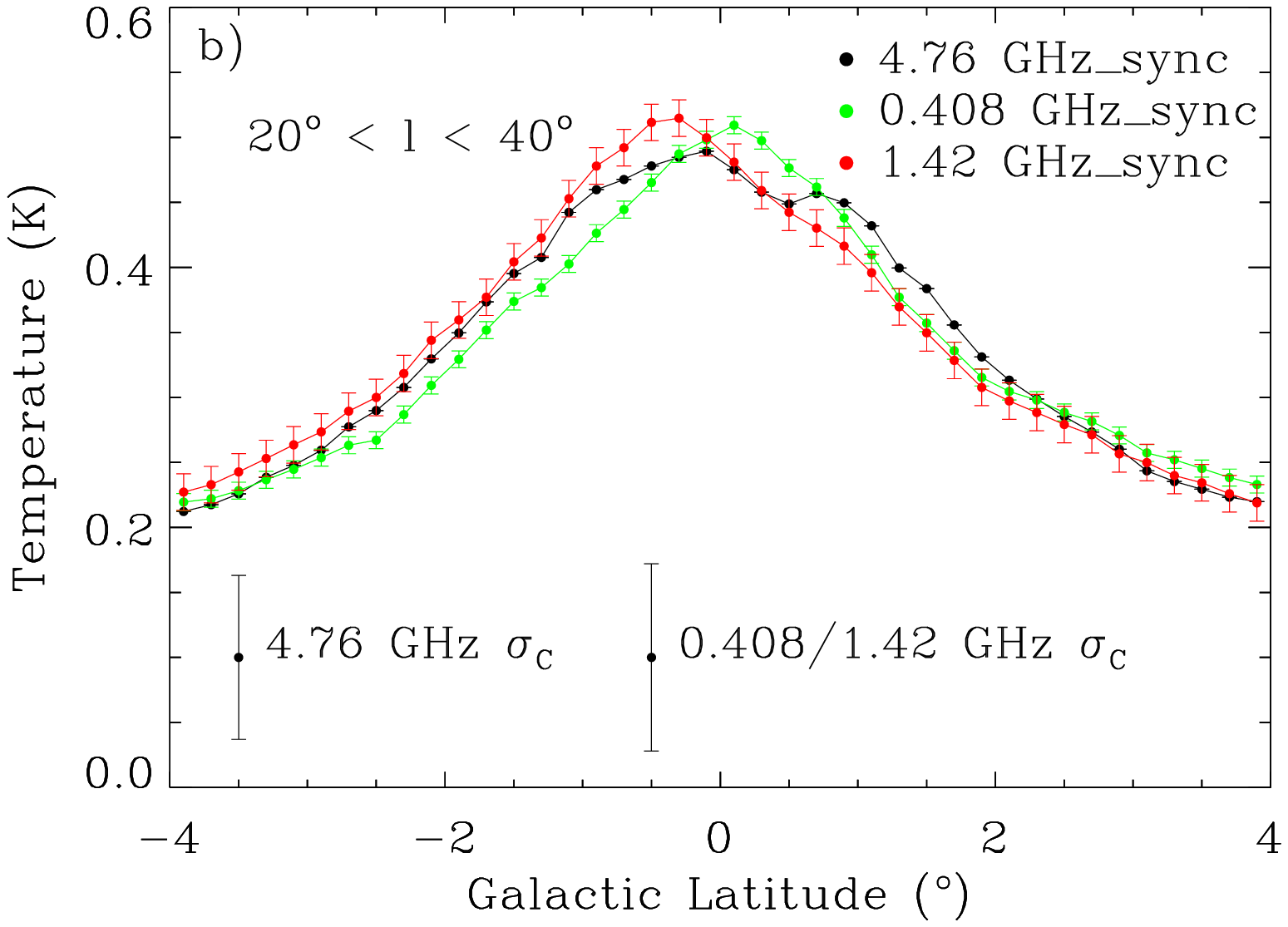}
  \caption{Latitude profiles of Galactic synchrotron emission 
     for $|b| < 4\degr$
    using averaged longitude data from the RRL free-free subtracted
    \protect{\protect\citet{Haslam}}, \protect{\protect\citet{Reich}} and C-BASS data, all scaled to
    4.76 GHz for the longitude range of a) 
21\degr\ -- 26\degr\ and b) 20\degr\ -- 40\degr\ (see Table~\ref{tab:syncIndAll}). The maps
    have been downgraded to {\tt{HEALPix}} $N_{\rm side} = 256$ and are
    smoothed to 1\degr\ resolution. The thermal noise is shown as error bars 
    on the data points while the calibration uncertainty (based on a 0.4\;K signal) are shown at the bottom of each panel. The broad profile expected for synchrotron emission is clearly visible.}
  \label{fig:syncRRL}
\end{figure}

Three synchrotron maps at 0.408\;GHz, 1.420\;GHz and \mbox{4.76\;GHz}
were made by subtracting the RRL free-free template from the
{\protect\citet{Haslam}}, {\protect\citet{Reich}} and \mbox{C-BASS}
maps. The RRL free-free template was scaled to the different
frequencies using a single power-law with $\beta = -2.1$. The
synchrotron maps, scaled to 4.76\,GHz (using $\beta$
  values given in Table~\ref{tab:syncIndAll}), are shown as latitude
profiles in Fig.~\ref{fig:syncRRL}. The best-fit synchrotron spectral
indices required to match the 0.408/1.420\,GHz latitude profiles with
their corresponding 4.76\,GHz values are shown in
Table~\ref{tab:syncIndAll}.  The data are shown at {\tt{HEALPix}}
$N_{\rm side} = 256$ so as to fully sample the latitude profile,
resulting in correlated pixel errors. The thermal noise is shown as
error bars while the average calibration errors (based on a 0.4\;K
signal) are shown in the legend. Fig.~\ref{fig:syncRRL}a averages over
the 21\degr\ to 26\degr\ longitude range while \;\!\ref{fig:syncRRL}b
averages over the entire 20\degr\ to 40\degr\ longitude range. As in
Fig.~\ref{fig:free}, the synchrotron curves are shown after the
subtraction of their zero-levels. Both Fig.~\ref{fig:syncRRL}a and
Fig.~\ref{fig:syncRRL}b display the broad peaks typically associated
with synchrotron emission. Although one average synchrotron spectral
index is clearly a good approximation across the $|b| = 4^{\circ}$
latitude range, 
small scale differences are seen between the curves. A full analysis of the 
intensity and polarization data sets will reveal which of these differences
are physically interesting.

These in- and off-plane results confirm a synchrotron spectral index of $-2.72 < \beta < -2.64$ between 0.408 and 4.76\;GHz for $ -10\degr < b < 10\degr$. Once again no significant spectral steepening of the synchrotron spectral index is observed and the values are in agreement with the expected index of $-2.7$. 

At 1.5\,GHz, diffuse Galactic emission has been shown to be roughly 30 per cent free-free and 70 per cent synchrotron emission \protect\citep{Platania98}. The ratio between free-free and synchrotron emission at 4.76\,GHz can now be determined using the derived latitude distributions of the 4.76\,GHz pure synchrotron map. Fig.~\ref{fig:comp} shows latitude profiles across $-4\degr < b < 4\degr$ for the total 4.76\,GHz emission as measured by C-BASS, the RRL free-free map scaled to 4.76\;GHz using $\beta = -2.1$ and the 4.76\,GHz pure synchrotron map. Extrapolating our results to 1.5\,GHz, the fraction of synchrotron emission present is $70 \pm 10$ per cent while at 4.76\,GHz it was found to be $53 \pm 8$ per cent for $ 20\degr < l < 40\degr$, $-4\degr < b < 4\degr$. 

 \begin{figure}
\centering 
 \includegraphics[width=0.48\textwidth]{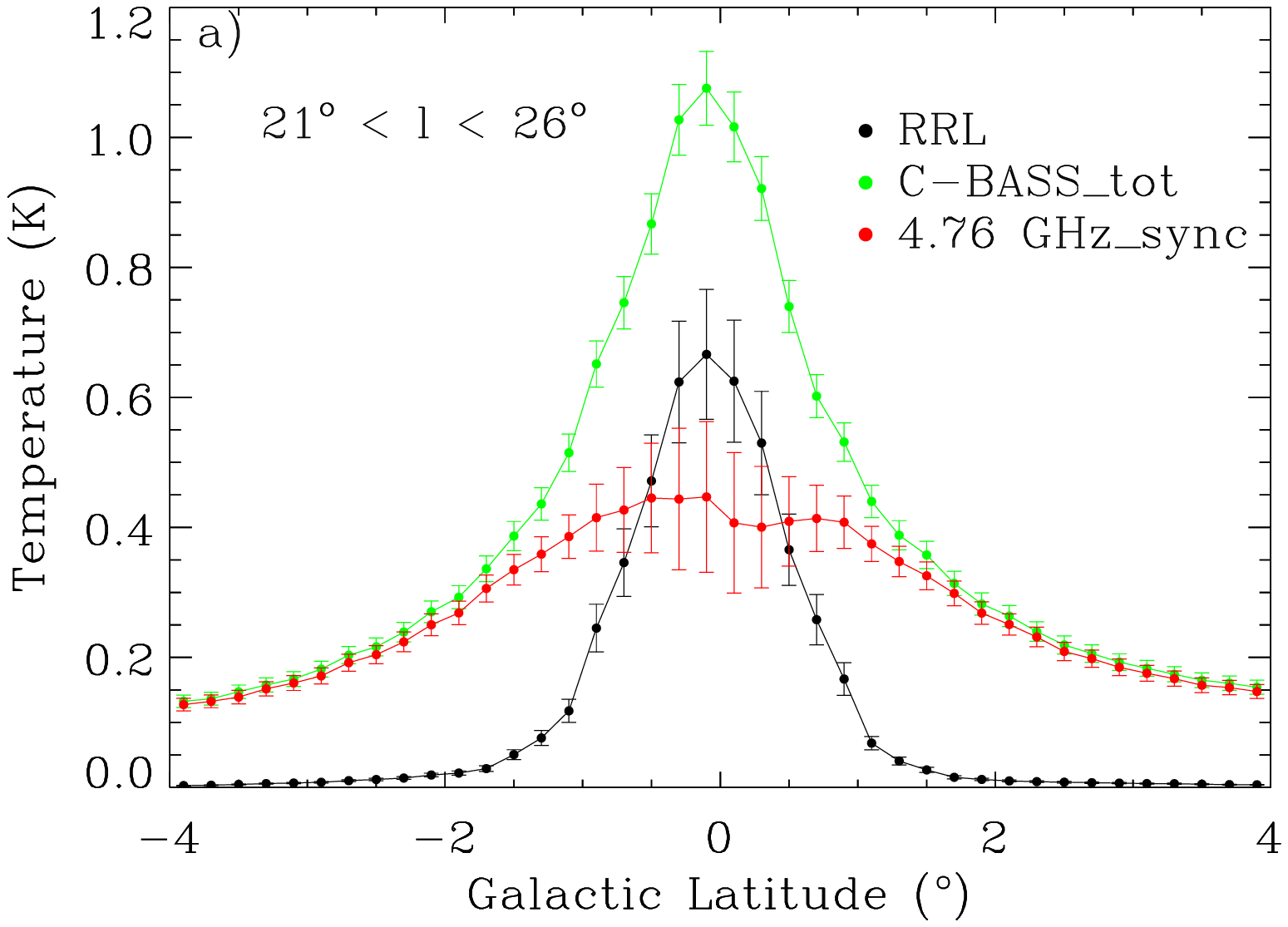} 
 \includegraphics[width=0.48\textwidth]{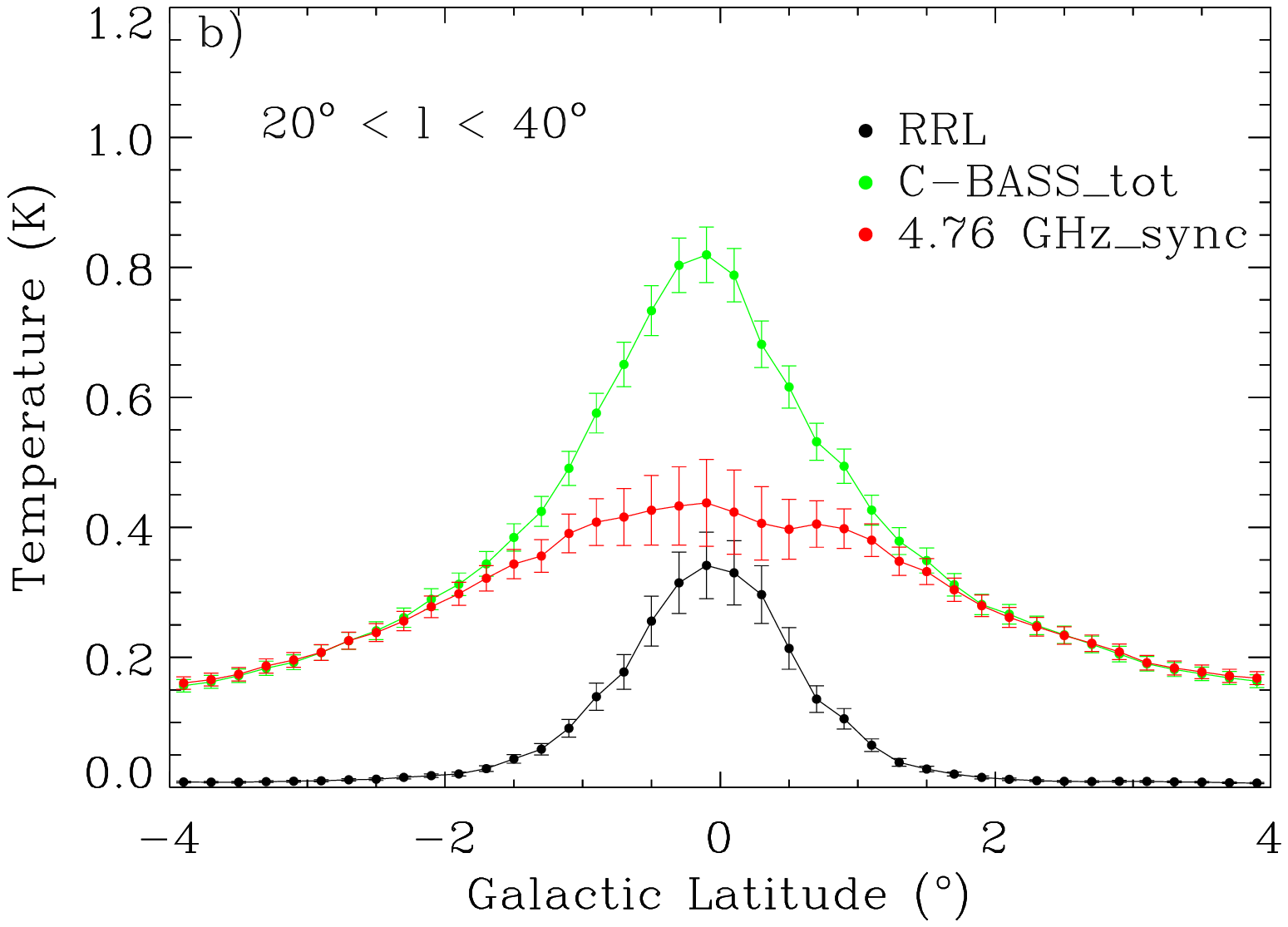}
 \caption{Total emission latitude distribution for $|b| < 4\degr$ using RRL and C-BASS data all at 4.76 GHz with the longitude data averaged between a) 21\degr\ and 26\degr\ and b) 20\degr\ and 40\degr. The maps have been downgraded to {\tt{HEALPix}} $N_{\rm side} = 256$ and are smoothed to 1\degr\ resolution.}
 \label{fig:comp}
\end{figure}

\section{Identification of Anomalous Emission at 22.8\;GH\lowercase{z}}

 At the \emph{WMAP} K-band frequency of 22.8\;GHz, diffuse emission in the Galactic plane is a combination of free-free, synchrotron and anomalous microwave emission (AME). The favoured emission mechanism for AME is electric dipole radiation from spinning dust grains {\protect \citep[e.g.][]{DL98, PlanckAME11, D96}}. AME has been readily identified in dark clouds and molecular clouds {\protect\citep{Vidal, Watson, ameCas, amiAME}}. However it is also present throughout the Galactic plane and adds to the complexity of component separation of CMB data {\protect\citep{PlanckAME11, PlanckAME13}}. For example, with only full-sky data at 1.4 and 22.8\;GHz, the presence of AME, which peaks in flux density at $\sim$ 30\,GHz {\protect\citep{PlanckAME11}}, will appear only as a flattening of the total-emission power-law. Here we use the preliminary C-BASS data to help characterise the synchrotron signal and so detect the presence of AME. The 4.76\;GHz C-BASS data provides a higher-frequency measure of the synchrotron emission, which is uncontaminated by AME, and can thus be used in combination with 22.8\;GHz data to separate the different spectral components. 
 
\begin{figure}
\centering 
\includegraphics[width=0.45\textwidth]{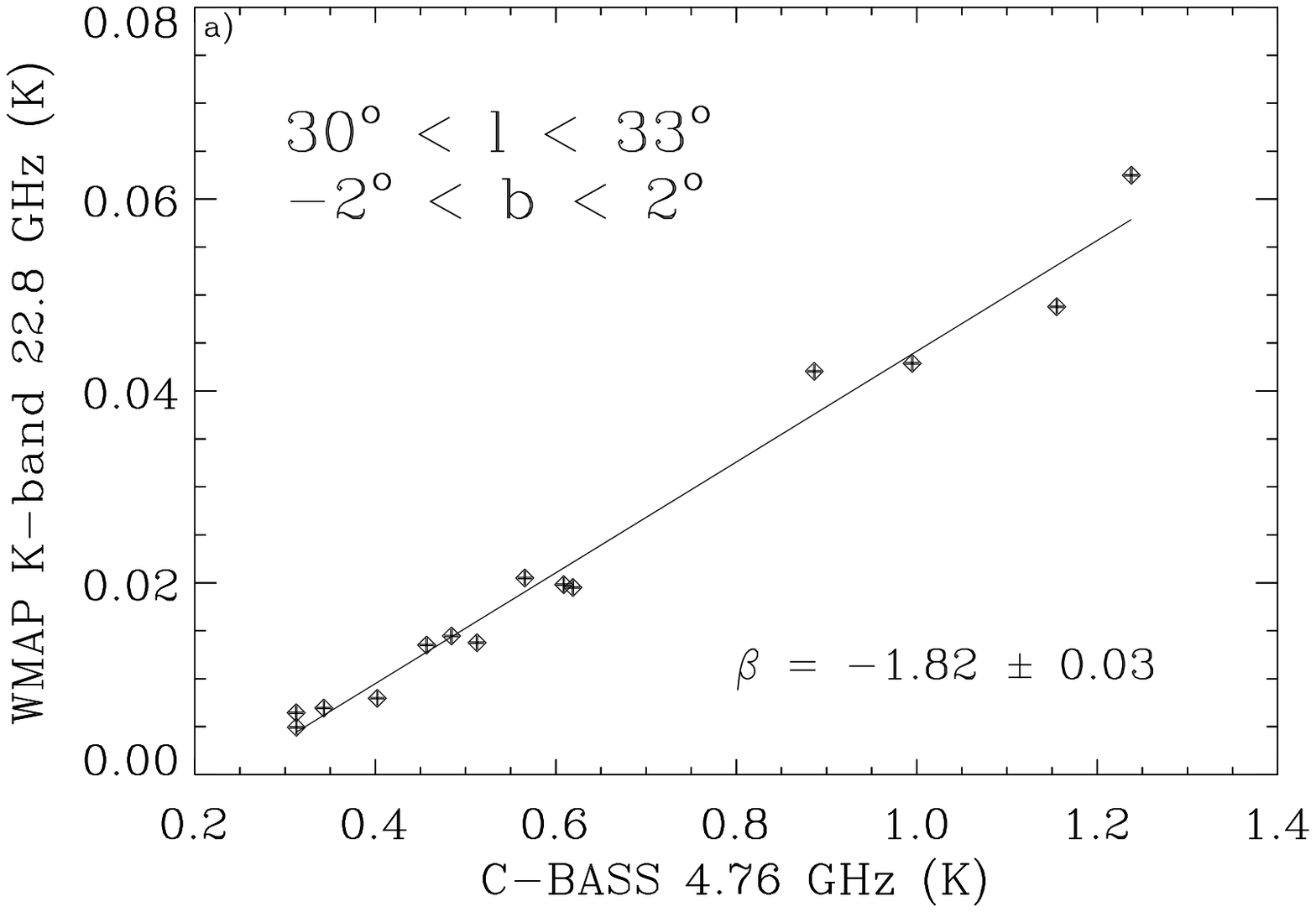} 
\includegraphics[width=0.45\textwidth]{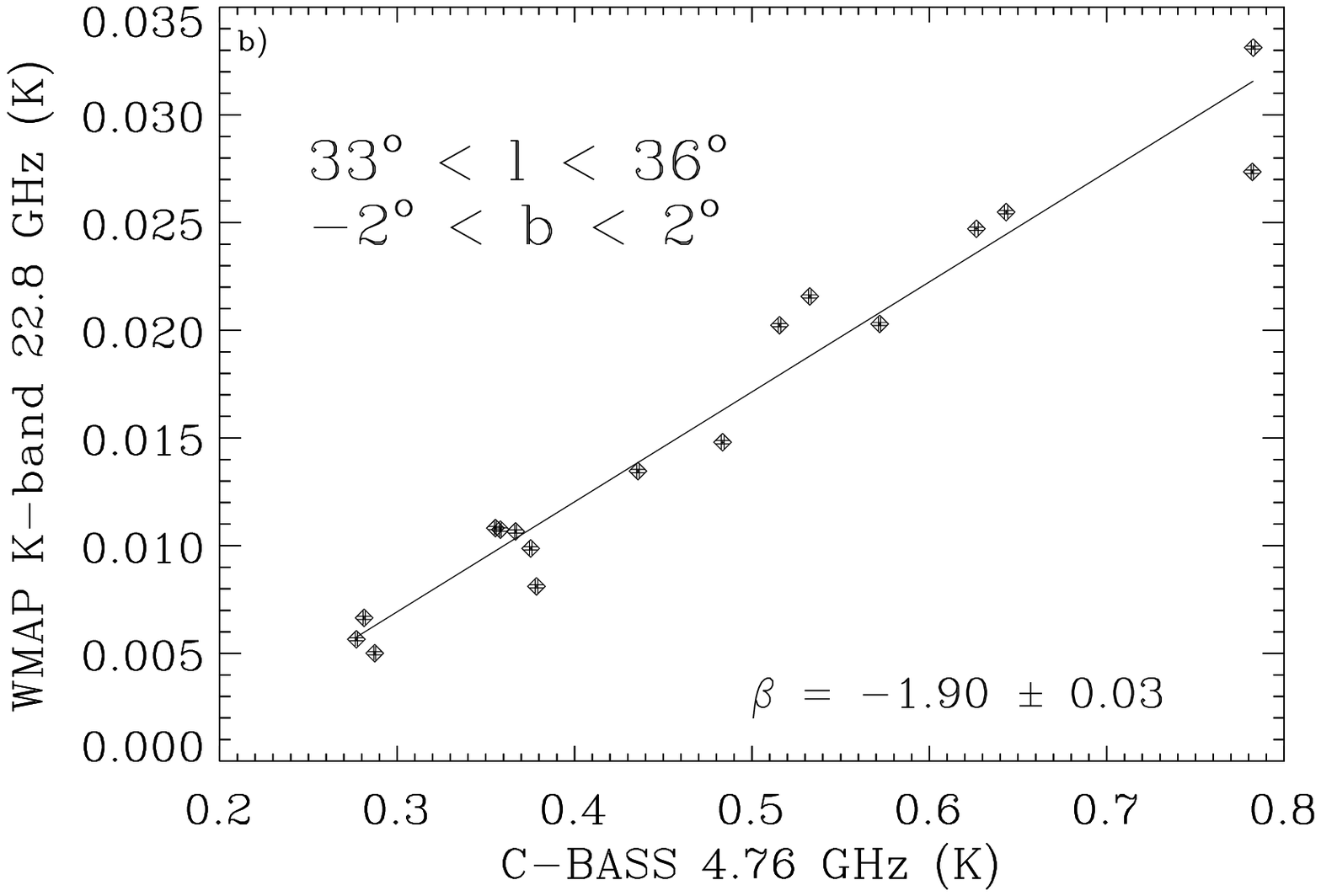}
\includegraphics[width=0.45\textwidth]{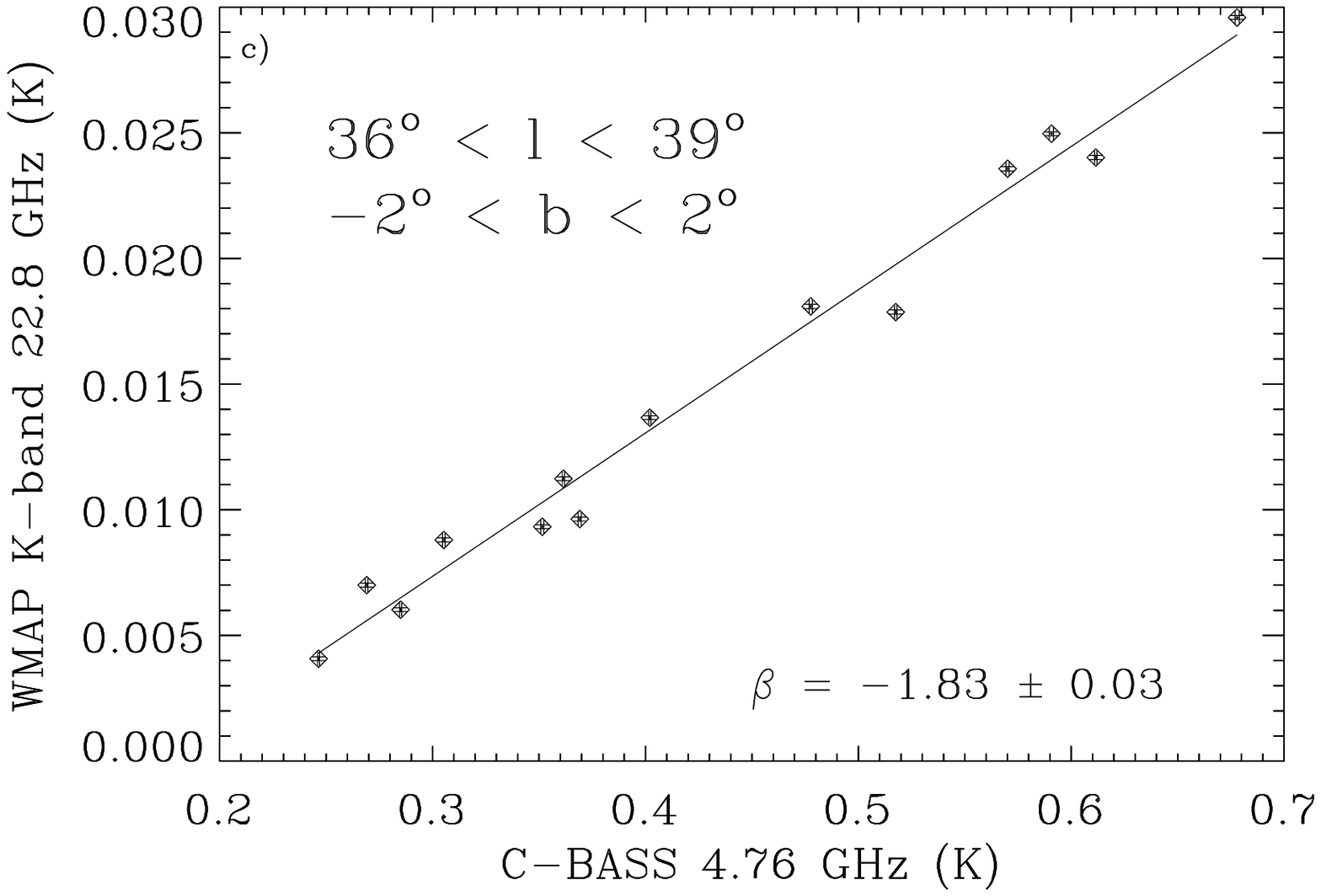} 
  \caption{$T$-$T$ plot between 4.76\,GHz and 22.8\,GHz for Galactic latitude $-2\degr < b < 2\degr$ and a) Galactic longitude $30\degr < l < 33\degr$, b) Galactic longitude $33\degr < l < 36\degr$ and c) Galactic longitude $36\degr < l < 39\degr$. All maps have been downgraded to $N_{\rm side} = 64$ and smoothed to 1\degr\ resolution. The presence of AME between 4.76 and 22.8\;GHz is implied by the flatter ($\beta > -$2.0) spectral indices.}
  \label{fig:anomTT} 
\end{figure}

\begin{table}
\centering
\caption{The Galactic plane synchrotron spectral indices as determined using synchrotron maps formed using the RRL free-free templates.}
 \begin{tabular}{  c  c  c  c }
\hline
Long. ($\degr)$ & $\nu$ range & $\beta$ \\
\hline \hline
 21 -- 26 & 0.408 -- 4.76 & $-2.59 \pm 0.08$  \\ 
 21 -- 26 & 1.420 -- 4.76 & $-2.69 \pm 0.17$ \\
 20 -- 40 & 0.408 -- 4.76 & $-2.56 \pm 0.07$  \\
 20 -- 40 & 1.420 -- 4.76 & $- 2.62 \pm 0.14$ \\
\hline
\end{tabular}
\label{tab:syncIndAll}
\end{table}

\subsection{Diffuse AME}

For this analysis the following three regions were chosen for their
combination of bright diffuse and compact regions: $30\degr < l <
33\degr$, $33\degr < l < 36\degr$ and $36\degr < l < 39\degr$ for
$-2\degr < b < 2\degr$. In this section we will use
  $T$-$T$ plots to probe the diffuse emission while in the next
  section will use spectral energy distributions (SEDs) to probe the
  compact regions.

At 22.8\,GHz the synchrotron contribution is negligible at low
Galactic latitudes, away from bright supernova remnants.  This is
apparent from scaling the 0.408\,GHz map to 22.8\,GHz using $\beta =
-2.7$ as the scaled {\protect\citet{Haslam}} signal is, on average, an
order of magnitude lower than the \emph{WMAP} intensity
signal. Therefore the spectral index determined from the
\emph{WMAP}--C-BASS $T$-$T$ plots is expected to be dominated by
free-free emission ($\beta \approx -2.1$). If spectral indices less
than $-2.1$ are found, the flattening would indicate the presence of
AME at 22.8\,GHz.

Fig.~\ref{fig:anomTT} shows $T$-$T$ plots between 4.76\,GHz and
22.8\,GHz for those three regions. The goodness-of-fit for all three
plots are fairly poor as spatial variations within the beam result in
outliers. The pixel uncertainties shown on the plots, are quite small
as, on the plane, the thermal noise is negligible. Therefore the
spatial variations within the beam dominate the formal spectral index
errors, which also take into account the survey calibration errors as
well as the plotted thermal errors.

Fig.~\ref{fig:anomTT} shows spectral indices of
$-1.82 \pm 0.03$, \mbox{$-1.90 \pm 0.03$} and $-1.83 \pm 0.03$ for
$30\degr < l < 33\degr$, \mbox{$33\degr < l < 36\degr$} and
$36\degr < l < 39\degr$. These indices are significantly flatter
than the expected free-free dominated total emission index of $\approx
-2.1$.  The percentage of the measured emission that can be
explained by AME emission can be calculated as follows:
\begin{equation}
\frac{{\rm{AME}}}{{\rm{Total}}} = 1 -  \frac{ (22.8/4.76)^{\beta_{e}} }{ (22.8/4.76)^{\beta_{\rm{tot}}}},
\end{equation}
where $\beta_{\rm{tot}}$ the total emission spectral index between
4.76\,GHz and 22.8\,GHz and $\beta_{e}$ is the expected spectral
index of $-2.12 \pm 0.02$. 


For the regions shown in Figs.~\ref{fig:anomTT}a,
\;\!\ref{fig:anomTT}b and \;\!\ref{fig:anomTT}c the percentage of the
total emission that can be explained as AME is calculated to be 37
$\pm 5$ per cent, \mbox{29 $\pm 6$ per cent} and \mbox{37 $\pm$ 8} per cent,
respectively. This is comparable to the 48 $\pm 5$ per cent AME to total emission ratio
that was found in the inner Galaxy using {\it Planck} data {\protect\citep{RodsSync}}.} This calculation provides a 7$\sigma$, 4$\sigma$ and 
4$\sigma$ detection of AME for the three $-2\degr\!\! < b <
2\degr\!\!$ and $30\degr < l < 39\degr$ subregions. These
result are summarised in Table~\ref{tab:ame} for a frequency of 22.8\,GHz. The table also 
contains the ratios of AME to free-free emission, which show the 
AME to be roughly half of the free-free magnitude.

\begin{table}
\setlength{\tabcolsep}{1.5mm}
\centering
\caption{The total emission spectral indices between 4.76 and 22.8\,GHz and the AME to total emission and AME to free-free emission ratios at 22.8\,GHz for \mbox{$-2\degr < b < 2\degr$}.}
 \begin{tabular}{ c  c  c c }
\hline
 Longitudes & $\beta_{\rm{tot}}$ & AME/tot & AME/ff \\
     ($\degr$)    &                        &(per cent)               &(per cent)         \\             
\hline \hline
30 -- 33 & $-1.82 \pm 0.03$ & 37 $\pm$ 5 (7$\sigma$) & 60 $\pm$ 13 (4$\sigma$)  \\ 
33 -- 36 & $-1.90 \pm 0.03$ & 29 $\pm$ 5 (4$\sigma$) & 41 $\pm$ 12 (3$\sigma$) \\
36 -- 39 & $-1.83 \pm 0.03$ & 37 $\pm$ 8 (4$\sigma$) & 58 $\pm$ 12 (4$\sigma$) \\
\hline
\end{tabular}
\label{tab:ame}
\end{table}

\subsection{Spectral Energy Distributions (SEDs)}
\label{sec:sed}

\begin{figure}
 \includegraphics[width=0.48\textwidth]{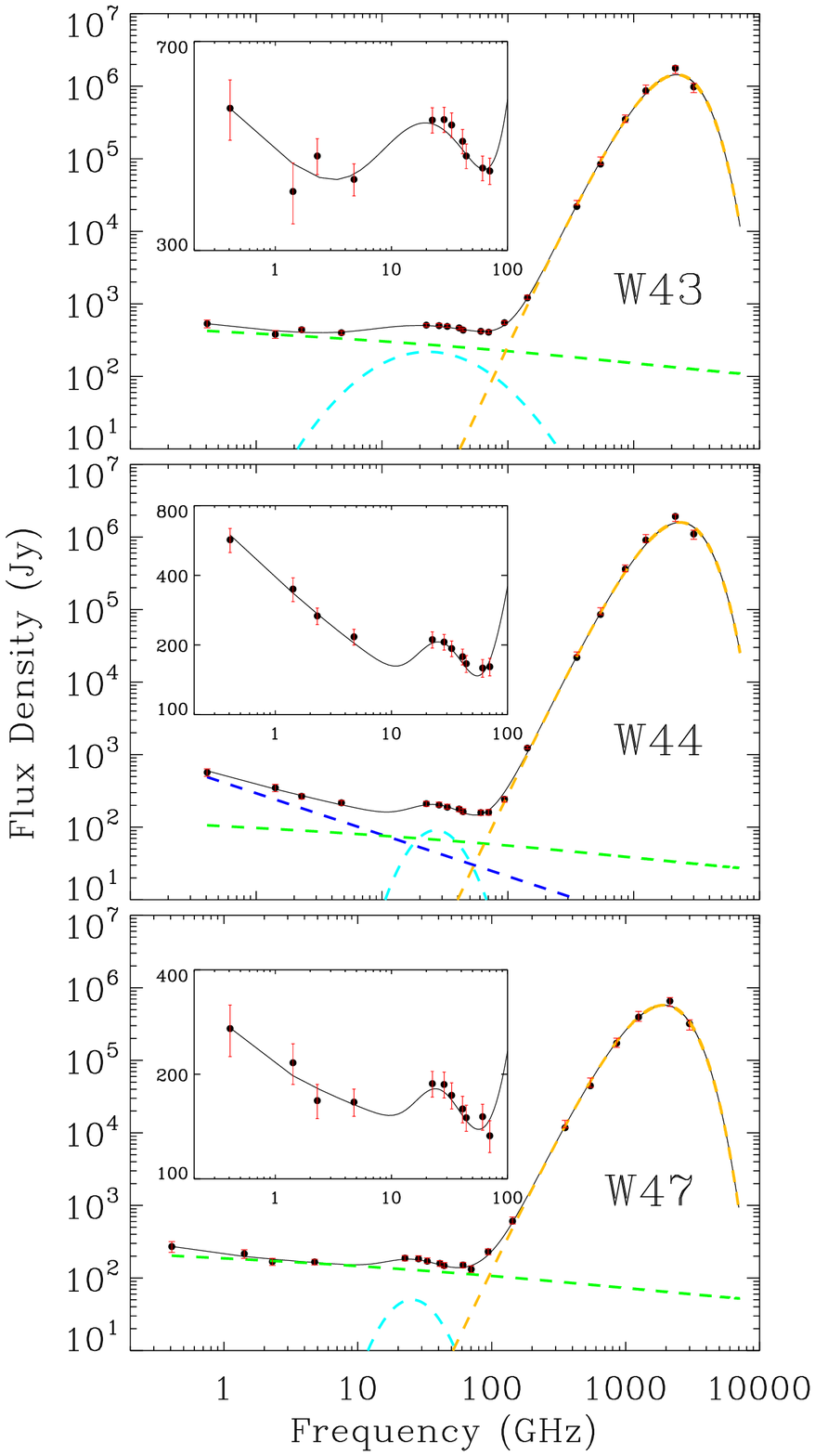}
   \caption{From top to bottom panel: SEDs for $(l,b)$ = (30$\degr\!\!.8$, $-$0$\degr\!\!.3$), $(l,b)$ = (34$\degr\!\!.8$, $-$0$\degr\!\!.5$) and $(l,b)$ = (37$\degr\!\!.8$, $-$0$\degr\!\!.2)$. Flux densities were determined via aperture photometry using the \protect{\protect\citet{Haslam}}, \protect{\protect\citet{Reich}}, \protect{\protect\citet{Jonas}}, C-BASS, \emph{WMAP}, \emph{Planck} and \emph{COBE}-DIRBE data. All the maps were smoothed to 1\degr\ resolution and downgraded to {\tt{HEALPix}} $N_{\rm side} = 256$. The solid black line represents the total emission model while the yellow dashed line represents the thermal dust emission model, the light blue dashed line represents AME, the dark blue dashed line represent synchrotron and the dashed green line represents free-free emission. All three sources clearly show the presence of AME in the form of a bump centered around 30\;GHz. }
   \label{fig:SED}
\end{figure}

Three prominent low-latitude extended sources, 
centred at  $(l,b) = (30\fdg 8,\, -0\fdg 3), (34\fdg 8,\,-0\fdg 5),
(37\fdg 8,\, -0\fdg 2)$, can be seen within Area 4 
demarcated on  Fig.~\ref{fig:Area}. 
 These are W43 (a
complex {H\sc{ii}} region), W44 (a SNR) and W47 (a
complex {H\sc{ii}} region), respectively.
To determine whether the presence of AME calculated
from the $T$-$T$ plots may be explained by excess emission due to these regions, 
we can calculate the SEDs for these individual 
complexes. 

Flux densities for complex extended sources are best determined
through aperture photometry. This technique measures the integrated
flux without the need to fit Gaussians. Aperture
photometry involves forming an inner aperture centered on the source
centre and an outer annulus over the source background.

SEDs of W43, W44 and W47 are shown in Fig.~\ref{fig:SED}. Zoomed-in plots of the AME spectrum are included within each SED to highlight the region of interest. To determine the source flux densities the maps were downgraded to {\tt{HEALPix}} $N_{\rm side} = 256$ when necessary. Aperture photometry was carried out and the inner aperture size was chosen to be 60\,arcmin in radius with the outer annulus spanning from 80\,arcmin to 100\,arcmin so as to match the analysis in {\protect\citet{PlanckAME13}}. The flux densities for W43, W44 and W47 are listed in Table~\ref{tab:fluxList}. The uncertainties are the quadrature addition of the calibration and the background subtraction uncertainties.    

\begin{table*}
 \catcode`@=\active
 \def@{  }
 \caption{The colour-corrected flux densities for the W43, W44 and W47 acquired using aperture photometry with a 60 arcmin inner aperture radius.}
 \begin{tabular}{ c  c  c c }
\hline
Frequency (GHz)&  & Flux Density (Jy) \\
  & W43 & W44 & W47  \\
  & (30$\degr\!\!.8$, $-$0$\degr\!\!.3$) & (34$\degr\!\!.8$, $-$0$\degr\!\!.5$) & (37$\degr\!\!.8$, $-$0$\degr\!\!.2)$ \\
  \hline \hline
0.408 & 534 $\pm$ 65@@@@@@@ & 570 $\pm$ 68@@@@@@ & 271 $\pm$ 46@@@@@@@  \\
1.420 & 381 $\pm$ 33@@@@@@@ & 349 $\pm$ 27@@@@@@ & 216 $\pm$ 22@@@@@@@@\\
2.3@@@ & 440 $\pm$ 32@@@@@@@ & 267 $\pm$ 22@@@@@@ & 168 $\pm$ 19@@@@@@@@\\
4.76@@@@ & 400 $\pm$ 26@@@@@@@ & 216 $\pm$ 17@@@@@@ & 166 $\pm$ 15@@@@@@@\\
22.8@@@@ & 509 $\pm$ 26@@@@@@@ &  211 $\pm$ 17@@@@@@ & 188 $\pm$ 16@@@@@@@\\
28.4@@@@ & 502  $\pm$ 26@@@@@@@ & 203 $\pm$ 16@@@@@@ & 184 $\pm$ 16@@@@@@@\\
33.0@@@@ & 490 $\pm$ 25@@@@@@@ & 189 $\pm$ 15@@@@@@ & 170 $\pm$ 15@@@@@@@\\
40.7@@@@ & 467 $\pm$ 23@@@@@@@ & 178 $\pm$ 14@@@@@@ & 159 $\pm$ 14@@@@@@@\\
44.1@@@@ & 435 $\pm$ 22@@@@@@@ & 164 $\pm$ 14@@@@@@ & 148 $\pm$ 13@@@@@@@\\
60.7@@@@ & 419 $\pm$ 21@@@@@@@ & 159 $\pm$ 14@@@@@@ & 151 $\pm$ 13 @@@@@@@\\
70.4@@@@ & 408 $\pm$ 22@@@@@@@ & 160 $\pm$ 14@@@@@@ & 132 $\pm$ 14@@@@@@@\\
93.5@@@@ & 551 $\pm$ 30@@@@@@@& 243 $\pm$ 21@@@@@@ & 231 $\pm$ 22@@@@@@@\\
143@@@@@ & 1213 $\pm$ 80@@@@@@@@ & 1231 $\pm$ 61@@@@@@@ & 610 $\pm$ 71@@@@@@@\\
353@@@@@ & $(2.20 \pm 0.17) \times 10^{4}$  & (2.19 $\pm$ 0.13) $\times 10^{4}$  & (1.18 $\pm$ 0.15) $\times 10^{4}$\\
545@@@@@ & (8.46 $\pm$ 1.1)$ \times 10^{4}$@@  & (8.55 $\pm$ 1.1) $\times 10^{4}$@ & (4.48 $ \pm$ 0.74)$\times 10^{4}$@@  \\
857@@@@@ & (3.49 $\pm$ 0.43)$ \times 10^{5}$@@ & (3.61 $\pm$ 0.41) $\times 10^{5}$ & (1.72 $\pm$ 0.26) $\times 10^{5}$@@\\
1249@@@@@ & (8.67 $\pm$ 1.2) $\times 10^{5}$@@ & (9.10 $\pm$ 1.2) $\times 10^{5}$ & (3.94 $\pm$ 0.63) $\times 10^{5}$@@\\
2141@@@@@ & (1.77 $\pm$ 0.21) $\times 10^{6}$ & (1.92 $\pm$ 0.21) $\times 10^{6}$ & (6.61 $\pm$ 0.091) $\times 10^{5}$\\
2997@@@@@ & (9.77 $\pm$ 1.4) $\times 10^{5}$@@@ & (1.10 $\pm$ 0.15) $\times 10^{6}$ & (3.20 $\pm$ 0.5) $\times 10^{5}$@@@\\
\hline
\end{tabular}
\label{tab:fluxList}
\end{table*}

The model used in Fig.~\ref{fig:SED} has nine free parameters giving 10 degrees of freedom and represents the following mixture of emission flux densities:
\begin{equation}
S = S_{{\rm{ff}}} + S_{{\rm{sync}}} + S_{{\rm{td}}}  + S_{{\rm{AME}}}, 
\end{equation} 
where the $S_{\rm{ff}}$ denotes the free-free contribution, $S_{\rm{sync}}$ the synchrotron contribution, $S_{\rm{td}}$ the thermal dust contributions and $S_{\rm{AME}}$ the AME contributions. A description of the spectral forms for each emission component can be found in \cite{PlanckAME13}. 

The synchrotron spectral behaviour was modelled as a power-law, with an amplitude and spectral index $\beta$. 
The free-free emission has a well-determined spectral shape with just the amplitude (Emission Measure, EM). 
The Gaunt factor ($g_{\rm{ff}}$) is required to take into account quantum mechanical contributions and the electron temperature is assumed to have a typical value of \mbox{7,000 K} {\protect\citep{ISM}}. 
The thermal dust was assumed to behave as a grey body with dust temperature and emissivity index. The amplitude is normalised using the 250 $\mu$m optical depth ($\tau_{250}$). 
The AME component was fitted as a parabola in log $S$--log $\nu$ {\protect\citep{Bonaldi}} 
where $m_{60}$ is a free parameter, which represents the slope of the parabola at 60\,GHz, $\nu_{\rm{peak}}$ is the AME peak frequency. This method requires no prior information on the physical properties of dust grains responsible for the AME.

Colour corrections were applied to the \emph{WMAP}, \emph{Planck} and \emph{COBE}-DIRBE data within the least squares fitting routine. This enables the model-fitted spectrum to be used as the source spectrum required for the colour correction. The \emph{WMAP} colour corrections are calculated by weighting the bandpass at a lower, central and upper limit, these limits and their weights are given in {\protect\citet{Wmap09}}. The IDL code used to compute the \emph{Planck} LFI and HFI  colour correction were taken from the {\it Planck} external tools repository\footnote{\url{http://externaltools.planck.fr/}} and the COBE colour correction tables are given on the LAMBDA website\footnote{\url{http://lambda.gsfc.nasa.gov/product/cobe}}. 

The W43 data are well fitted by the total emission model with a
reduced $\chi^{2}$ of 1.0. The AME peak, the largest difference between 
the total emission model and the free-free + synchrotron + thermal 
dust emission model, is identified at the
4.4$\sigma$ level. The large  uncertainty on the $m_{60}$ parameter 
(see Table~\ref{tab:SEDparams})
comes from the fact that few of the measurements are characterised 
by this parameter. The W44 data are the least well fitted by the total
emission model with a reduced $\chi^{2}$ of 2.6. This is
due to a slightly poorer fit between the three \emph{COBE}-DIRBE
points and the grey body model of thermal
dust. If the reduced $\chi^{2}$ is recalculated using the 
full data set to determine the model parameters but leaving
out the \emph{COBE}-DIRBE points to determine the goodness-of-fit,
a value of 1.9 is found. This is assuming the full ten degrees 
of freedom. The AME peak is identified at the 3.1$\sigma$ level. 
The W47 data are well fitted by the total emission model with a reduced
$\chi^{2}$ of 0.6. The AME peak is identified at the 2.5$\sigma$
level. It should be noted that the $EM$ fitted values are only
effective $EM$ values for the 1 degree radius inner aperture.

The analysis was repeated with the 4.76\;GHz C-BASS point excluded
from the fits to quantify the impact of the additional constraint. Table~\ref{tab:SEDparams} shows the fitted parameters for
W43, W44 and W47 with and without the C-BASS data point included
in the fit. It can be seen that the lack of 4.76\;GHz data increases
the uncertainty on the synchrotron spectral index, from 0.08 to 0.09,
for W44 and the
uncertainty associated with the free-free $EM$ of W43. The 0.01
variation in the W44 synchrotron spectral index may in fact represent an 
actual value shift after the inclusion of additional synchrotron 
information. The AME flux decreases by 17 per cent for W43, 
increases by 7 per cent for W44 and 
8 per cent for W47 and the uncertainties on the dust temperatures of 
W44 and W47 triple. The AME peak
frequencies for W44 and W47 are seemingly unaffected by the
C-BASS data point, while the W43 AME peak frequency has a lower 
uncertainty when the C-BASS point is included. 

The level of improvement seen when the C-BASS point
  is included in the fit is clearly strongly dependent on the region
  under investigation. In these three regions, it turns out that the
  C-BASS data point is in excellent agreement with the fitted spectral
  model and therefore it has a minimal impact. Furthermore, these
  regions benefit from the use of the 2.3\,GHz data point. Also, from
Fig.~\ref{fig:SED} it can be seen that the W43 AME model parabola is
far wider than the W44 or W47 parabola. The width of the W43 AME peak
even when the C-BASS data point is included in the fit suggests that
the models are still not well enough
constrained. Data between 4.76 and 22.8\;GHz would be of great use for
this spectral fit, for example the 11, 13, 17 and 19\;GHz measurements
of the QUIJOTE experiment {\protect\citep{lopez}}.

\begin{table*}
\caption{The W43, W44 and W47 fitted total emission parameters with and without the C-BASS data point. The parameters, from top to bottom, represent the reduced chi squared, the spectral index of synchrotron emission, the dust temperature, the spectral index of thermal dust emission, the spinning dust slope parameter, the peak frequency of spinning dust emission, the emission measure, the optical depth at 250 $\mu$m and the AME peak flux density.}
\centering
\protect
    \begin{tabular}{ l |  c c | c c | c c }
\hline
 & W43 & & W44 & & W47 \\
 \hline
Parameter & With & Without & With  & Without & With & Without \\ 
\hline \hline
Reduced $\chi^{2}$ & 1.0 & 1.0 & 2.6 & 2.8 & 0.7 & 0.8\\
$\beta_{\rm{sync}}$ & -- & -- & $-2.57\pm 0.08$ & $-2.56 \pm 0.09$ & -- & -- \\
$T_{\rm{d}}$ (K) & $23.3 \pm 0.5$ & $23.3 \pm 0.6$ & $25.2 \pm 0.5 $ & $25.3 \pm 1.5$& 19.9 $\pm$ 0.4 & 21.3 $\pm$ 1.5  \\
$\beta_{\rm{thermal}}$ & $1.78 \pm 0.06$ & $1.77 \pm 0.05$ & $1.69 \pm 0.05$ & $1.69 \pm 0.07$ & $1.91 \pm 0.07$ & $1.80 \pm 0.07$\\
$m_{60}$ & $1 \pm 0.2$ & 2 $\pm$ 1& $4 \pm 2$ & $5 \pm 2$ &  $5 \pm 4$ & $7 \pm 6$  \\
$\nu_{\rm{peak}}$ (GHz) & $23 \pm 3$ & $ 22 \pm 9 $ &  $27 \pm 3$ & $28 \pm 2$ &  $25 \pm 5$ &$26 \pm 6$ \\
EM & $97 \pm 8$ & 77 $\pm$ 7 & $24 \pm 1$ & 24 $\pm$ 2 & $46 \pm 2$ & $49 \pm 3$ \\
$\tau_{250}$ & $0.011 \pm 0.0006$ & $0.011 \pm 0.002$ & $0.009 \pm 0.0007$ & 0.007 $\pm$ 0.0008& $0.009 \pm 0.0009$ & $0.007 \pm 0.0006$ \\
AME peak flux (Jy) & 218 $\pm$ 29 & 186 $\pm$ 41 & 90 $\pm$ 13 & 96 $\pm$ 14 & 50 $\pm$ 11 & 54 $\pm$ 4 \\
\hline
\end{tabular}
\label{tab:SEDparams}
\end{table*}


\subsubsection{Contribution from UCHII}

The Planck Collaboration have previously noted the possibility of AME
within W43 {\protect\citep{PlanckAME13}}. For their analysis they used
aperture photometry to determine the flux density values and they
measured a $\sigma_{\rm{AME}}$ of 8.6, however they assign a 36 per
cent contribution to the total emission to emission from nearby
ultra-compact H{\sc ii} regions. Therefore we will
  now consider the possible contribution of ultra-compact H{\sc ii}
  regions to the detection of diffuse AME.

Ultra compact H{\sc ii} (UCHII) regions are dense areas of ionised hydrogen, which form around hot stars. They are defined as possessing an $EM > 10^{6}$ cm$^{-6}$ pc, an electron temperature ($T_{b}$) of $\sim 10^{4}$\,K and electron density of \mbox{$10^{5}$ -- $10^{6}$\,cm$^{-3}$} {\protect\citep{ISM}}. At low frequencies, typically under 15\,GHz {\protect\citep{UCH11}}, UCHII regions may be optically thick with a flux spectral index of $\beta = +4$, moving into the optically thin regime ($\beta = -2.1$) at frequencies higher than 15\,GHz:
\begin{equation}
T_{b} =  T_{e} \left(1 - e^{-\tau}  \right)
\end{equation} 
where $\tau \ll$ 1 indicates the optically thin regime and $\tau \gg$ 1 indicates the optically thick regime.

If a UCHII region, optically thick at 5\;GHz, were to be positioned
within one of the source regions this would result in what would
appear to be excess emission at higher frequencies where the UCHII
region becomes optically thin. To ensure that the excess emission seen
around 30\,GHz in the W43, W44 and W47 SEDs is due to AME, the sources must be checked
for nearby optically thick UCHII regions.

In the \protect\citet{PlanckAME13} paper the method of {\protect\cite{CliveUCH11}} and {\protect\cite{Wood}} was employed. This method uses the fact that UCHII regions generally possess IRAS colours (brightness ratios) of $\log_{10} \left(\frac{S_{60}}{S_{12}} \right) \geq 1.30$ and $\log_{10} \left(\frac{S_{25}}{S_{12}} \right) \geq 0.57$. 

Recently a new catalogue of UCHII regions at 5\,GHz has become
available via the CORNISH (Co-Ordinated Radio `N' Infrared Survey for
High mass star formation) VLA survey. The CORNISH survey covers the
$10\degr < l < 65\degr, |b| < 1\degr$ region of the Galactic plane at
a resolution of 1.5\,arcsec and r.m.s. noise of 0.4\;mJy per beam
{\protect\citep{Cornish}}. Using the on-line
catalogue\footnote{\url{http://cornish.leeds.ac.uk/public/catalogue.php}},
several possible UCHII regions were found within a 0\fdg 5 radius of
W43 and W47 and their flux densities and optical depth values are
shown in Table~\ref{tab:uch11}. The flux densities and FWHMs are
listed in the CORNISH catalogue and used to calculate the brightness
temperature, which in turn are used to calculate the optical depth at
5\;GHz.

\begin{table}
 \catcode`@=\active
 \def@{  }
 \caption{The 5\;GHz flux density, FWHM, brightness temperature and 5\;GHz optical depth values of the UCHII regions found within 0\fdg 5 of W43, W44 and W47 using the CORNISH 5\;GHz catalogue.}
\renewcommand{\tabcolsep}{3pt}
\begin{tabular}{ c  c  c c c}
\hline
Source & $S$ (mJy)& FWHM ($^{\prime \prime}$) & $T_{b}$ (K) & $\tau_{5}$ \\
\hline \hline
& & W43 & & \\ 
G030.8662$+$00.1143 & 325@@@ & 3.1 & 1667@@ & 0.20 \\
 G030.7532$-$00.0511 & 302@@@ & 3.3 & 1363@@ & 0.15\\
G031.1596$+$00.0448  & 24@@ & 2.4 & 206@ & 0.02\\
 G030.7197$-$00.0829 & 969@@@ & 4.6 & 2250 & 0.25 \\
G030.6881$-$00.0718  & 467@@@ & 10.8 & 195 & 0.02\\
G030.7661$-$00.0348  & 88@@@ & 3.3 & 403 & 0.04 \\
 G030.5887$-$00.0428 & 92@@@ & 1.8 & 1410 & 0.15 \\
G030.5353$-$00.0204  & 710@@@ & 6.3 & 875 & 0.09 \\
G030.9581$+$00.0869  & 26@@ & 3.2 & 123 & 0.01\\
G031.0595$+$00.0922  & 12@@ & 2.0 & 142 & 0.01 \\
 G031.2420$-$00.1106 & 296@@@ & 7.8 & 238 & 0.02 \\
 G031.2448$-$00.1132 & 37@@ & 3.1 & 186 & 0.02\\
G031.2435$-$00.1103  & 353@@@ & 2.7 & 2386 & 0.27\\
  G031.1590$+$00.0465 & 7 & 2.0 & 84 & 0.01 \\
\hline
& & W44 & & \\
G035.0524$-$00.5177 & 68@@ & 3.7 & 246@ & 0.02\\
\hline
 & & W47 & & \\
G037.5457$-$00.1120 & 407 @@@& 8.2 &297@ & 0.03\\
 G037.7347$-$00.1128 &  16@@ & 1.8 & 253@ & 0.03\\
 G037.8731$-$00.3996 &  2561@@@@ & 8.9 & 1574@@@ & 0.17\\
 G037.8683$-$00.6008 &  210@@@ & 5.2 & 383@ & 0.04\\
 G037.9723$-$00.0965 & 21@ & 2.5 & 170 & 0.02\\
\hline
\end{tabular}
\renewcommand{\tabcolsep}{6pt}
\label{tab:uch11}
\end{table}

The opacity values for these regions range from 0.01 to 0.2 and so it can be
seen that none of the UCHII sources near W43, W44 and
W47 are optically thick at 5\,GHz. Using a free-free spectral index
of $-2.12$ we see that their combined fluxes would contribute 0.013\;Jy
at 30\;GHz for W43, 0.008\;Jy at 30\;GHz for W44 and 0.048\;Jy at 30\,GHz for W47. As their combined flux densities are a small fraction ($\sim 1$ per cent) of the total flux density measured in a $1^{\circ}$ beam the
excess emission at frequencies 20 -- 60\;GHz associated with W43, W44 and W47 cannot be solely due
to UCHII regions and is therefore attributed to AME.

\section{Conclusion}

We have presented intensity data from a preliminary \mbox{C-BASS} map made from two months of northern receiver data. This map has a high signal-to-noise ratio with negligible contributions from instrumental or confusion noise. The overall intensity calibration is shown to be good to $\approx 5$ per cent. The C-BASS data allow us to study the Galactic radio emission from synchrotron and free-free radiation.

For $20\degr < l < 40\degr$, $4\degr < |b| < 10\degr $ we have determined a synchrotron spectral index of $-2.69 < \beta <  -2.65$ between 0.408\,GHz and 4.76\,GHz and $-2.72 < \beta <  -2.64$ between 1.420\,GHz and 4.76\,GHz, confirming that the synchrotron spectral index in this region is consistent with a straight power law with $\beta = -2.7$. After subtracting a free-free template based on radio recombination lines from the Galactic plane region ($-4\degr < b < 4\degr, 20\degr < l < 40\degr$), we measured synchrotron spectral indices of $-2.56 \pm 0.07$ between \mbox{0.408 GHz} and \mbox{4.76 GHz} and \mbox{$-2.62 \pm 0.14$} between 1.420\,GHz and 4.76\,GHz. Again, consistent with $\beta = -2.7$. The synchrotron fraction of the total emission at 4.76\,GHz is $53 \pm 8$ per cent. Additionally, this separation also provides latitude distributions for Galactic synchrotron and free-free components. 

Combining C-BASS data with the higher frequency data of \emph{WMAP}, \emph{Planck} and \emph{COBE}-DIRBE we have detected diffuse AME within $-2\degr\!\! < b < 2\degr\!\!,~30\degr < l < 40\degr$ at the 4--7$\sigma$ level. This region contains the compact sources W43, W44 and W47 and so we used SEDs to show that AME can be associated with each of these regions, the largest detection being 4.4$\sigma$ for W43. We detected diffuse AME within the ISM at a $\sim 4\sigma$ level, specifically at $|l| = 30\degr$--39$\degr$, $|b|<2\degr$. AME accounts for $\sim 30$ per cent of the total emission at 22.8 GHz and $\sim 50$ per cent of the free-free intensity.

The C-BASS  final data will have a thermal noise level significantly lower than the data shown here, better control of potential systematics and better 1/\emph{f} removal in both polarization and intensity across the full sky.  With the full survey data, AME regions like the Perseus and Ophiuchi molecular clouds will also benefit from a 5\,GHz data point to further constrain their emission spectra. The expected C-BASS sensitivity in polarization is 0.1\;mK per beam. With the polarization data in hand we will be able to trace the synchrotron polarization amplitude and direction unmodified by Faraday rotation, thus improving our knowledge of Galactic foregrounds.

\section*{Acknowledgments}

The C-BASS project is a collaboration between Caltech/JPL in the US,
Oxford and Manchester Universities in the UK, and Rhodes University
and the Hartebeesthoek Radio Astronomy Observatory in South Africa. It
is funded by the NSF (AST-0607857, AST-1010024, and AST-1212217), the University of Oxford, the
Royal Society, and the other participating institutions. We would like to
thank Russ Keeney for technical help at OVRO. We thank the Xilinx University Programme for their donation of FPGAs to this project.
MI acknowledges studentship funding from an EU Marie-Curie IRG grant (no.~230946) awarded to CD. CD acknowledges an STFC Advanced Fellowship and an ERC Starting Grant (no.~307209). JZ acknowledges support from the European Research Council is the form of a Starting Grant with number 240672. PGF acknowledges support from Leverhulme, STFC, BIPAC and the Oxford Martin School. We thank Andy Strong for helpful discussions and the anonymous referee who provided useful suggestions to help clarify several aspects in the paper. We make use of the IDL astronomy library and {\tt{HEALPix}} \protect\citep{Gorski}.

\bibliographystyle{mn2e}
\bibliography{DiffuseGalacticEmission.bib}

\appendix

\bsp

\label{lastpage}

\end{document}